\pdfoutput=1

\documentclass[
]{frontiersSCNS-nologo} 

\usepackage[T1]{fontenc} 
\input{glyphtounicode} \pdfgentounicode=1
\usepackage[utf8]{inputenx}

\newcommand*{\oggi}{3 May 2018}
\newcommand*{\propertitle}{Inferring health conditions\\ from fMRI-graph data}
\newcommand*{\pdftitle}{Inferring health conditions from fMRI-graph data}
\newcommand*{\headtitle}{Health conditions from fMRI data}
\newcommand*{\pdfauthor}{P.G.L. Porta Mana, C. Bachmann, A. Morrison}

\usepackage[onehalfspacing]{setspace}

\usepackage{textcomp}
\usepackage{amsmath}
\usepackage{mathtools}
\usepackage{empheq}

\usepackage{framed}
\usepackage{amssymb}
\usepackage{amsxtra}
\usepackage[Euler]{upgreek}

\usepackage[main=british,french,italian,german,swedish,latin,esperanto]{babel}
\newcommand*{\langfrench}{\foreignlanguage{french}}

\newcommand*{\langitalian}{\foreignlanguage{italian}}

\usepackage[autostyle=false,autopunct=false,english=british]{csquotes}
\setquotestyle{american}

\usepackage[shortlabels,inline]{enumitem}
\SetEnumitemKey{para}{itemindent=2\parindent,leftmargin=0pt,listparindent=\parindent,parsep=0pt,itemsep=\topsep}
\setlist[enumerate,2]{label=\alph*.}
\setlist[enumerate]{leftmargin=2\parindent,labelsep=0.75\parindent,topsep=\smallskipamount,itemsep=\topsep}
\setlist[itemize,2]{label=--}
\setlist[itemize]{leftmargin=2\parindent,labelsep=0.75\parindent,topsep=\smallskipamount,itemsep=\topsep}

\usepackage{mathdots}

\usepackage[usenames]{xcolor}
\definecolor{mypurpleblue}{RGB}{68,119,170}
\definecolor{myblue}{RGB}{102,204,238}
\definecolor{mygreen}{RGB}{34,136,51}
\definecolor{myyellow}{RGB}{204,187,68}
\definecolor{myred}{RGB}{238,102,119}
\definecolor{myredpurple}{RGB}{170,51,119}
\definecolor{mygrey}{RGB}{187,187,187}
\definecolor{shadecolor}{gray}{0.9}

\usepackage{bm}
\usepackage{microtype}

\newcommand*{\citey}{\citeyearpar}
\providecommand{\href}[2]{#2}

\newcommand*{\citein}[2][]{\textnormal{\citet[#1]{#2}}
}
\newcommand*{\citebi}[2][]{\citet[#1]{#2}
}
\newcommand*{\subtitleproc}[1]{}
\newcommand*{\chapb}{}

\newcommand*{\arxiveprint}[1]{
\texttt{\urlalt{https://arxiv.org/abs/#1}{arXiv:\hspace{0pt}#1}}%
}

\newcommand*{\osfeprint}[1]{%
\texttt{\urlalt{https://doi.org/10.17605/osf.io/#1}{doi:10.17605/osf.io/#1}}%
}

\usepackage{graphicx}
\pdfmapline{+dummy-space <dummy-space.pfb}\pdfinterwordspaceon%

\PassOptionsToPackage{hyphens}{url}\usepackage{hyperref}
\usepackage[depth=4]{bookmark}
\hypersetup{colorlinks=true,bookmarksnumbered,pdfborder={0 0 0.25},citebordercolor={0.2 0.1333 0.5333},
citecolor=mypurpleblue,linkbordercolor={0.2666 0.4666 0.6666},
linkcolor=myredpurple,urlbordercolor={0.9333 0.4 0.4666},
urlcolor=mygreen,breaklinks=true,pdftitle={\pdftitle},pdfauthor={\pdfauthor}}
\providecommand*{\urlalt}{\href}

\usepackage[british]{datetime2}
\DTMnewdatestyle{mydate}%
{
}
\DTMsetdatestyle{mydate}

\newcommand*{\chap}{ch.} 
\newcommand*{\chaps}{chs} 
\newcommand*{\sect}{\S} 
\newcommand*{\sects}{\S\S} 
\newcommand*{\fig}{fig.} 
\newcommand*{\eqn}{eq.}
\newcommand*{\eqns}{eqs}
\newcommand*{\ie}{i.e.}
\newcommand*{\eg}{e.g.}
\newcommand*{\etc}{etc.}

\newcommand*{\cf}{{cf.}}

\newcommand*{\amp}{\&}

\definecolor{notecolour}{RGB}{34,136,51}

\newcounter{mynotecounter}

\newcommand*{\widebar}[1]{{\mkern1.5mu\skew{2}\overline{\mkern-1.5mu#1\mkern-1.5mu}\mkern 1.5mu}}

\newcommand*{\zerob}[1]{\makebox[0pt][c]{#1}}
\newcommand*{\defd}{\coloneqq}
\newcommand*{\pu}{\uppi}
\newcommand*{\di}{\mathrm{d}}
\newcommand*{\p}{\mathrm{P}}
\newcommand*{\pf}{\mathrm{p}}
  
\renewcommand*{\|}{\mathpunct{|}}
\newcommand*{\bigcond}{\mathpunct{\big|\ }}

\newcommand*{\Land}{\mathbin{\ \land\ }}
\newcommand*{\Cond}{\mathpunct{\,|\ }}

\let\oldsum\sum
\let\oldprod\prod
\renewcommand*{\sum}{\mathop{\textstyle\oldsum}}
\renewcommand*{\prod}{\mathop{\textstyle\oldprod}}

\newcommand*{\tland}{\mathop{\textstyle\bigwedge}\nolimits}
\newcommand*{\tsum}{\mathop{\textstyle\sum}\nolimits}
\DeclarePairedDelimiter\set{\{}{\}}
\newcommand*{\T}{^\intercal}
\DeclarePairedDelimiter\clcl{[}{]}
\DeclarePairedDelimiter\clop{[}{[}

\DeclarePairedDelimiter\opop{]}{[}

\DeclareMathSymbol{\de}{\mathalpha}{letters}{"40}
\newcommand*{\E}{\mathrm{E}}
\DeclarePairedDelimiter\expp{(}{)}
\newcommand*{\expe}{\E\expp}

\newcommand*{\yH}{C}
\newcommand*{\yh}{c}
\newcommand*{\yhu}{\textsc{h}}
\newcommand*{\yhd}{\textsc{s}}
\newcommand*{\yD}{D}
\newcommand*{\yF}{F}

\newcommand*{\yf}{\bm{f}}
\newcommand*{\yxx}{f}
\newcommand*{\yx}{\bm{\yxx}}

\newcommand*{\ylm}{\lambda}
\newcommand*{\ylmm}{\bm{\ylm}}
\newcommand*{\yls}{\varLambda}
\newcommand*{\ylss}{\bm{\yls}}

\newcommand*{\yp}{q}
\newcommand*{\yph}{q_{\yhu}}
\newcommand*{\yps}{q_{\yhd}}
\newcommand*{\ypph}{\Hat{p}_{\yhu}}
\newcommand*{\ypps}{\Hat{p}_{\yhd}}
\newcommand*{\ypth}{p_{\yhu}}
\newcommand*{\ypts}{p_{\yhd}}

\newcommand*{\ypc}{\Hat{\yp}}

\newcommand*{\yn}{n}
\newcommand*{\ynh}{\yn_{\yhu}}
\newcommand*{\yns}{\yn_{\yhd}}
\newcommand*{\yd}{d}
\newcommand*{\yI}{I}
\newcommand*{\yM}{M}
\newcommand*{\yMl}{\yM_l}

\newcommand*{\ytr}{l}
\newcommand*{\yT}{\textsc{t}}
\newcommand*{\ynT}{\textsc{d}}
\newcommand*{\yu}{u}
\newcommand*{\yuT}{\yu_{\textsc{t}}}
\newcommand*{\yuD}{\yu_{\textsc{d}}}
\newcommand*{\yudh}{\yu_{\textsc{d}\|\textsc{h}}}
\newcommand*{\yuds}{\yu_{\textsc{d}\|\textsc{s}}}
\newcommand*{\yuth}{\yu_{\textsc{t}\|\textsc{h}}}
\newcommand*{\yuts}{\yu_{\textsc{t}\|\textsc{s}}}

\newcommand*{\av}{\widebar}
\DeclareMathOperator{\Cov}{Cov}

\newcommand*{\yth}{\bm{\theta}}
\newcommand*{\ythh}{\yth_{\yhu}}
\newcommand*{\yths}{\yth_{\yhd}}
\newcommand*{\yL}{L}
\newcommand*{\dnormal}{\mathrm{N}}

\DeclareMathOperator{\diwishart}{Wishart^{-1}}
\DeclareMathOperator{\dstudentt}{t}

\newcommand*{\ymu}{\bm{\delta}}
\newcommand*{\yka}{\kappa}
\newcommand*{\ynu}{\nu}

\newcommand*{\yLa}{\bm{\varDelta}}
\newcommand*{\ymuo}{\ymu_0}
\newcommand*{\ykao}{\yka_0}
\newcommand*{\ynuo}{\ynu_0}
\newcommand*{\yLao}{\yLa_0}
\newcommand*{\id}{\bm{I}}
\DeclareMathOperator{\tr}{tr}

\newcommand*{\tom}{\text{T}}
\newcommand*{\joe}{\text{J}}

\def\keyFont{\fontsize{8}{11}\helveticabold }
\def\firstAuthorLast{Porta Mana, Bachmann, Morrison}
\def\Authors{ P.G.L. Porta Mana\,$^{1}$, C. Bachmann\,$^{2,*}$,A. Morrison\,$^{2,3,4}$}
\def\Address{$^{1}$Kavli Institute for Systems Neuroscience, NTNU Norwegian
  University of Science and Technology, Norway\\
$^{2}$Institute of Neuroscience and Medicine (INM-6) and Institute for Advanced Simulation (IAS-6) and JARA BRAIN    Institute I, J\"ulich Research Centre, J\"ulich, Germany\\
$^{3}$Simulation Laboratory Neuroscience, Inst. for Advanced Simulation, JARA, J\"ulich Research Centre and JARA, J\"ulich, Germany\\
$^{4}$Institute of Cognitive Neuroscience, Faculty of Psychology, Ruhr-University  Bochum, Germany}
\def\corrAuthor{C. Bachmann}
\def\corrEmail{c.bachmann@fz-juelich.de}

\begin{document}
\onecolumn
\firstpage{1}

\title[\headtitle]{\propertitle} 

\author[\firstAuthorLast ]{\Authors} 
\address{} 
\correspondance{} 

\extraAuth{}

\maketitle

\DTMsetdatestyle{mydate}
{30 January 2018; updated \oggi}

\begin{abstract}

\section{}

Automated classification methods for disease diagnosis are currently in the
limelight, especially for imaging data. Classification does not fully meet
a clinician's needs, however: in order to combine the results of multiple
tests and decide on a course of treatment, a clinician needs the likelihood
of a given health condition rather than binary classification yielded by
such methods. We illustrate how likelihoods can be derived step by step
from first principles and approximations, and how they can be assessed and
selected, using fMRI data from a publicly available data set containing
schizophrenic and healthy control subjects, as a working example. We start
from the basic assumption of partial exchangeability, and then the notion
of sufficient statistics and the \enquote{method of translation}
(Edgeworth, 1898) combined with conjugate priors. This method can be used
to construct a likelihood that can be used to compare different
data-reduction algorithms. Despite the simplifications and possibly
unrealistic assumptions used to illustrate the method, we obtain
classification results comparable to previous, more realistic studies about
schizophrenia, whilst yielding likelihoods that can naturally be combined
with the results of other diagnostic tests.

\tiny
 \keyFont{ \section{Keywords:} disease diagnosis,  decision theory, sufficient statistics,
   exchangeability,  parametric statistical model,
   schizophrenia, fMRI, Bayesian probability theory } 
\end{abstract}


\section{Introduction}
\label{intro}

A 29-year-old man seeks medical advice because he finds himself in a very
confused state. The clinician, after listening to the complaints of the
patient, identifies some diseases that would account for the symptoms.
However, the presentation is not clear cut, and treatment for some of the
potential conditions have significant side effects. To come to a decision
on the best course of action, the clinician decides to perform the
differential diagnosis in a mathematically sound manner
\citep{soxetal1988_r2013}, first assigning an initial probability for the
patient's being healthy or having each of the potential conditions, taking
into account age, sex, familial factors, symptoms, a psychological
evaluation, the incidence of the disease, and similar prior information:
\begin{equation}\label{eq:pre-test_prob_intro}
\p(\text{health condition}\Cond\text{prior info}).
\end{equation}

Then she orders one or more diagnostic tests to make a better informed
assessment of the probabilities of the considered diseases. Among these
tests she orders a structural and functional magnetic-resonance imaging
(MRI) scan. The advantage of MRI lies in the non-invasive monitoring of
brain structure and activity; the structural image (sMRI) is used to
exclude morphological changes in the brain such as tumours, and the
functional imaging (fMRI) can provide information about changes in brain
activity.

With the results of the tests and of the sMRI and fMRI, the clinician
updates her initial or prior probability to a \enquote{post-test} or
posterior probability based on the results, according to Bayes's theorem:
\begin{multline}\label{eq:bayes_theorem_intro}
    \overbrace{\vphantom{\bigg\{}\p(\text{health condition}\Cond
      \text{results of all tests}\Land
      \text{prior info})}^{\zerob{\scriptsize\emph{post-test probability}}}
    \mathrel{\ \propto\ }
    \overbrace{\vphantom{\bigg\{}\p(\text{health condition} \Cond
      \text{prior info})}^{\zerob{\scriptsize\emph{initial probability}}}
  \mathbin{\ \times\ }{}    
    \\[\jot]
\text{\scriptsize\emph{likelihoods}}\left\{\;  \begin{aligned}
 \p(\text{result of first test} \Cond \text{health condition} \Land
  \text{prior info})
  &\mathbin{\ \times\ }{}\\
  \p(\text{result of second test} \Cond \text{health condition} \Land
  \text{prior info})
  &\mathbin{\ \times\ }{}
\\
{}\dotsb &\mathbin{\ \times\ }{}\\
 \p(\text{result of sMRI} \Cond \text{health condition} \Land
   \text{prior info})&
 \mathbin{\ \times\ }{}\\
 \p(\text{result of fMRI} \Cond \text{health condition} \Land
  \text{prior info})&
  \end{aligned}\right.
\end{multline}
where \enquote{$\land$} denotes logical conjunction (\enquote{and}), and we
have reasonably assumed that the result of each test does not depend on
those of the other tests, \ie\ that their likelihoods are independent
\citetext{\citealp[\sect~4.2]{jaynes1994_r2003};
  \citealp[\sect~4.7]{soxetal1988_r2013}}.

In the update formula above, the initial probability is assessed by the
clinician. To calculate the post-test probability she needs the
probabilities for each test result conditional on the health condition,
either \enquote{healthy} or \enquote{presumptive disease}. These probabilities
are called the \emph{likelihoods for the health condition} in view of each
test. The term \enquote{likelihood} has its standard technical meaning in
the present work: the probability of a proposition $A$ given $B$ is
$\p(A\|B)$, while the likelihood of $A$ in view of $B$ is $\p(B\|A)$, \ie,
$A$ appears in the conditional
\citetext{\citealp[\sect~4.1]{jaynes1994_r2003};
  \citealp[\sect~6.1]{good1950}}. A proposition can have high probability
but low likelihood and vice versa. Probabilities, not likelihoods, are what
we base our decisions upon.

The final, post-test probability is necessary to the clinician to decide
upon a course of action \citetext{\citealp[\chap~6]{soxetal1988_r2013};
  \citealp{goodman1999}; \citealp[\sect~5.7]{murphy2012}}; for example, to
treat the patient according to one or another specific treatment, to
dismiss him, or to order more tests. To make such decision the clinician
will combine her post-test probabilities for the health conditions with a
utility table (a reminder of decision theory is given in
\sect~\ref{sec:decision_theory}).

In the following we assume that one of the presumptive diseases
the clinician has in her mind is schizophrenia. Although currently MRI does not play a role in a diagnosis of schizophrenia, there are substantial efforts to develop such analyses for this purpose \citep{silvaetal2014}. In this work, we focus on the diagnosis of this particular disease simply as a concrete worked example, to demonstrate how results of a diagnostic test, in this case
results from function MRI imaging, can be incorporated in the diagnostic process in a principled fashion.

In short, we address the question: \textbf{how can we assign a
  numerical value to the likelihood}
\begin{equation}
    \p(\text{fMRI result} \Cond \text{health condition} \Land
    \text{prior info}).
\label{eq:test_prob_intro}
\end{equation}
\textbf{of each health condition ('healthy' or 'schizophrenic') in view of the fMRI result?} We will propose an answer that can be applied for any brain disease. 

To this end it is useful to mark out some features of the
approach presented so far:
\begin{enumerate}[wide,label=\Roman*.]
\item\label{item:test_combination}\textbf{Modularity.} The update
  formula~\eqref{eq:bayes_theorem_intro} combines evidence from different
  tests, and this combination does not need to be done at once. The
  clinician can multiply her initial probability by the likelihood from the
  first test, normalize, and thus obtain a \enquote{post-first-test}
  probability. Later she can multiply this probability by the likelihood
  from the second test, normalize, and thus obtain a post-second-test
  probability; and so on with any number of other tests, a number that the
  clinician needs not fix in advance. She can therefore store the value of
  the likelihood from the fMRI result, to later combine it with new
  likelihoods from future tests to form a new, better-informed post-test
  probability.

\item\label{item:no_classif}\textbf{Decision-theoretic character.} The
  clinician's final goal is not simply a healthy/schizophrenic
  classification, but a \emph{decision} upon a course of action about the
  patient \citetext{\citealp[\chaps~6, 7]{soxetal1988_r2013};
    \citealp[\chaps~13, 14]{jaynes1994_r2003};
    \citealp{raiffaetal1961_r2000}}. This distinction is important: for
  example, a treatment without contraindications might be recommended even
  if there is only a 10\% probability that the disease is present; or a
  dangerous treatment might be recommended only if there is a 90\%
  probability that the disease is present.

  The modularity of the present approach extends to the decision stage,
  because the post-test probability can be used with different decisions
  and utilities, which can also be updated later on. For example, after
  beginning a treatment the clinician happens to read about a new kind of
  treatment, having new benefits and contraindications. Using the post-test
  probabilities she already has, she may re-evaluate her decision using an
  updated utility table that includes the new treatment.

\item\label{item:learning}\textbf{Incomplete knowledge.} In general, we lack a complete
  biological understanding of the relation between brain activity and the
  health condition under study. In this case, the likelihoods can only be
  assessed by relying on examples of known \emph{health condition--fMRI
    data} pairs, usually called a training or calibrating dataset. Moreover, this
  training dataset is often very small.
  
\item\label{item:large_dataspace}\textbf{High dimensionality.} The 
  fMRI data are positive-valued vectors with
  $10^7$--$10^8$ components or more \citep{lindquist2008}. This high
  dimension impacts the calculation of likelihoods and probabilities.
\end{enumerate}

\bigskip

The first two points above are great advantages of the present approach,
and also the reasons why it cannot be based on  machine learning algorithms for deterministic classification; such methods give the
clinician a dichotomous, \enquote{healthy/schizophrenic} answer, with no
associated uncertainty. This answer cannot be used by the clinician to
weigh the benefits and risks of different courses of action, the assessment
of which needs the probabilities of the health conditions. Probabilistic
algorithms, on the other hand, are not flexible for combining evidence:
they give a probability for the health condition, not a likelihood; and
only the latter can be combined with the likelihoods from other tests, or
stored for later reuse and combination.

We therefore approach the question of assigning the
likelihoods~\eqref{eq:test_prob_intro} by means of the probability
calculus, the same calculus from which \eqn~\eqref{eq:bayes_theorem_intro}
is derived. What we will do is in essence no different from current
Bayesian statistical analyses and modelling; but we would like to emphasize
some aspects of this modelling that are usually left in the background. The
probability calculus can be regarded as the extension of formal logic
(truth calculus) to plausible
inference
\citep{jeffreys1939_r2003,jaynes1994_r2003,hailperin1996}, a view also
supported in medicine \citep{greenland1998,maclure1998,goodman1999}, which
has been proven with increasing rigour by Koopman
\citey{koopman1940,koopman1940b,koopman1941}, Cox
\citey{cox1946,cox1961,cox1979}, P\'olya
\citey{polya1949,polya1954b_r1968}, and many others
\citep{horvitzetal1986,paris1994_r2006,halpern1999b,snow1998,dupreetal2009,tereninetal2015_r2017}.
The derivation of a probability proceeds much like an
\enquote{axioms~$\rightarrow$ logic rules~$\rightarrow$ theorem} derivation
in formal logic: one starts from the probabilities of some propositions,
and by applying the probability rules, arrives at the probability of the
desired proposition, \eqn~\eqref{eq:test_prob_intro} in our case.

We will show this procedure step by step in the case of our
problem, in order to expose where assumptions and approximations enter
the derivation. These may be improved by other researchers, or replaced by
different ones when the method is applied to a different problem. Our
discussion is inspired by Mosteller \amp\ Wallace's
\citep{mostelleretal1963} brilliant, thoughtful analysis of a statistically
similar problem in a very different context.

The approach we follow deals naturally with the four points listed above.
The small size of the training dataset, point~\ref{item:learning} above, is
not an issue because the probability calculus allows for training datasets
of any size. In fact, the calculus allows us to continuously update our
inferences given new training data, making our inferences more and more
precise and less likely to be affected by outliers.

The unmanageable size of our data space, point~\ref{item:large_dataspace}
above, will force us to make auxiliary assumptions that will translate into
the choice of a reduced data space, discussed in
\sect~\ref{sec:data_reduction}, and into the use
of parametric statistical models, discussed in \sect~\ref{sec:sufficiency}.
Regarding the latter, we will emphasize that assumptions about relevant and
irrelevant information in our data may translate into mathematical
statistical models. It is often difficult to relate biophysical
considerations about quantities measured in the brain to the shape of a
probability distribution, especially in multidimensional quantities. The
notion of \emph{sufficient statistics} \citetext{\citealp{dawid2013};
  \citealp[\chap~4]{bernardoetal1994_r2000};
  \citealp[\sect~5.5]{lindley1965b_r2008};
  \citealp{diaconisetal1981,cifarellietal1982,lauritzen1982_r1988,kallenberg2005}},
discussed in \sect~\ref{sec:suff_stat}, is a helpful bridge between
biophysical considerations and probability distributions. The idea is that
it may be easier for us to conceive a connection between biophysical
considerations and some special statistics of our measurements, than
between biophysical considerations and an abstract multidimensional
distribution function. This \enquote{translation} is powerful, because if
one finds the assumptions about relevance or irrelevance of some data
unreasonable, one can then make different assumptions, resulting in a
different statistical model.

Models inspired by sufficient statistics -- especially their comparison and
selection -- can nevertheless be computationally demanding owing to the
multidimensional integrals in their formulae, even when these are addressed by modern numerical methods such as
Monte Carlo \citetext{\citealp[\chap~IV]{mackay1995_r2003};
  \citealp[\chaps~23--24]{murphy2012}}. In the present study we shall
use analytically tractable statistical models, but availing ourselves of
Edgeworth's \enquote{Method of Translation}
\citetext{\citeyear{edgeworth1898}; \citealp{johnson1949};
  \citealp{mead1965}}: the simple but potentially very fertile idea of
transforming a quantity into a normally distributed one, discussed in
\sect~\ref{sec:generalized_normals}.

The possible combined choices of reduction of the data space, of sufficient
statistics, and of transformations into normal variable, lead to a variety
of possible models and likelihoods to be used by the clinician. Which is
the \enquote{best} one? We discuss several criteria for choice in
\sect~\ref{sec:model_comparison}, settling on one based on expected
utility. We also briefly discuss the remarkable observation that common Bayesian
criteria based on weight of evidence and Bayes factors
\citetext{\citealp[\chaps~V, VI, A]{jeffreys1939_r2003};
  \citealp{good1950,mackay1992,kassetal1995}} for the fMRI data gives
results opposite to those of the expected-utility criterion.

In this article, we will calculate the likelihoods for the health
conditions, \eqn~\eqref{eq:test_prob_intro}, and assess the models for so
doing, in the following steps. First, in \sect~\ref{sec:data_acquisition},
we briefly discuss schizophrenia and the use of fMRI to diagnose it,
introducing a concrete dataset of fMRI data for schizophrenic and healthy
patients. We then show that a simple and natural assumption, called
\emph{exchangeability}, would lead to a unique value of the
likelihoods~\eqref{eq:test_prob_intro} if the training dataset were large
enough (\sect~\ref{sec:exchangeability}). However, with a small training
dataset we must face two problems: unmanageably large dimensions of the
data space, and the need to specify prior beliefs, also involving functions
on infinite-dimensional spaces.
  
To solve the first problem, in \sect~\ref{sec:data_reduction} we assume
that information adequate for our health inference can be found in a
reduced data space of the fMRI, which we construct from time correlations
between groups of voxels. To solve the second problem we introduce
parametric statistical models in \sect~\ref{sec:sufficiency} using the
notions of \emph{sufficient statistics} and of transformation into normal
variables, mentioned above. We discuss how these models learn from the data
and select three models as possible candidates. We then consider several
criteria to select one of the three models against our data, as an example,
and discuss how a more realistic assessment could be made in a real
application (\sect~\ref{sec:model_comparison} ). We conclude with a
discussion (\sect~\ref{sec:discussion}) on how the choice of sufficient
statistics and prior probabilities could be improved, and on the relation to
machine-learning methods.

Our statistical terminology and notation follow ISO standards
\citep{iso1993_r2009,iso2006}.

\section{Results}
\label{sec:results}

\subsection{Selection of clinical use case and fMRI-data acquisition}
\label{sec:data_acquisition}

Schizophrenia is a psychiatric disorder that comprises various symptoms
that are categorized into positive (e.g. hallucinations), negative (e.g.
loss of motivation) and cognitive (e.g. memory impairment) disease
patterns. A common disease cause for all these widespread symptoms is still
unknown. Functional magnetic resonance imaging (fMRI) has been used to gain
insight into modifications in functional connectivity in this disease. In the resting state, functional connectivity is measured either
by asking the subject to fulfil a certain task or at rest, instructing the
subject to think about nothing specific but not fall asleep. In this
condition, both increased and decreased
functional connectivity have been reported in the default mode network,
although the hyperactivity seems to be reported more often \citep{Hu2017}.
Moreover, widespread connectivity changes in the dorsal attention network
and the executive control network have been detected \citep{Woodward2011,Yu2012}.

Beyond these individual sub-networks, many studies have found profound
changes in macroscopic brain structures, e.g. a shrinkage of whole brain
and ventricular volume, reduced gray matter in frontal, temporal cortex and
thalamus, and changes in white matter volume in frontal and temporal cortex
\citep{Shenton2010,Wright2009,Wright2010}. Since both gray matter loss and
white matter changes are found, it is reasonable to conclude that not only
the intrinsic activity of single areas is modified, but also the interplay
of different brain areas, in particular in frontal and temporal cortex. It
has been argued that these alterations in long range connectivity are
responsible for a range of disease symptoms that are not attributable to
single areas \citep{Friston1995}. Taking this disconnect hypothesis as a
starting point, we can reach the working hypothesis that these changes are
also reflected in the functional activity of the brain, and that fMRI
images can be used to distinguish schizophrenic from healthy patients. We
therefore conclude that schizophrenia is an appropriate condition to
demonstrate our approach.

We requested data of schizophrenic and healthy patients from
Schizconnect\footnote{http://schizconnect.org/},
a virtual database for public schizophrenia neuroimaging data. In our
request we asked for resting state T2*-weighted functional (rfMRI) and
T1-weighted structural magnet resonance images (MRI) from patients
participating in the COBRE study either with no known disorder or diagnosed
as schizophrenic according to the Diagnostic and Statistical Manual of
Mental Disorders (DSM) IV, excluding schizoaffective disorders. In the
COBRE study, the voluntary and informed participation of the subjects was
ensured by the institutional guidelines at the University of New Mexico
Human Research Protections Office. The ensuing dataset comprised 91 healthy
patients and 74 schizophrenic patients. Out of these we randomly selected
54 healthy and 49 schizophrenic subjects, to permit demonstration our
method on a small dataset with unequal group size. A detailed description
on the exact experimental design and the MRI scanning is provided by
\cite{Cetin2014}.

\subsection{Calculation of probabilities: exchangeability}
\label{sec:exchangeability}
Let us describe our context more precisely and set up some mathematical
notation. We have:
\begin{itemize}[para]
\item A number of possible health conditions, in our example healthy
  ($\yhu$) and schizophrenic ($\yhd$). The variable $\yh$ denotes health
  condition.

\item A space of possible fMRI data. They are vectors with $10^7$--$10^8$
  or more positive components \citep{lindquist2008}. The variable $\yf$
  denotes an fMRI result.

\item A set of $\yn$ patients, labelled in some way, the variable $i$
  denoting their labels. These labels may reflect information about the times
  the patients were examined, or about their geographical location. This
  possibility is important in the considerations to follow. In our study
  $\yn = 104$.

\item Knowledge of the health condition and of the fMRI result of each
  patient. Let us use the propositions
  \begin{equation}
    \label{eq:propositions_H_F}
    \begin{aligned}
      \yH_i^{\yh} &\defd \text{\enquote{Patient $i$ has
                    health condition $\yh$}},
      \\
      \yF_i^{\yf} &\defd \text{\enquote{The fMRI of
    patient $i$ gives $\yf$}},
    \end{aligned}
  \end{equation}
  the latter to be understood within a very small interval
  $(\yf,\yf+\di\yf)$. 
  In our study we have
  $\ynh=55$ healthy and $\yns=49$ schizophrenic patients.

  For brevity we denote by $\yH^{\yh}$ the conjunction of the propositions
  $\yH_i^{\yh}$ for all patients having health condition $\yh$, \ie\ our
  knowledge about which patients have that health condition; and
  analogously for $\yF^{\yh}$. By $\yD^{\yh}$ we denote all data about
  patients with health condition $\yh$; by $\yD$ we denote all our data.

\item An imaginary patient, labelled \enquote{$0$}, whose fMRI result $\yf$
  is known, but whose health condition is not.

\item Other pre-test information, denoted by $\yI$; for example the
  clinician's initial diagnosis of the health condition of patient $0$, and
  the results of any other diagnostic tests.

\item The probabilities $(\ypph,\ypps)$ for the health condition of patient
  $0$, conditional on the pre-test information, including the results from
  other tests. We call these \emph{pre-test probabilities}. Note that they
  may differ from the \emph{initial} probabilities of
  \eqns~\eqref{eq:pre-test_prob_intro}--\eqref{eq:bayes_theorem_intro},
  because they may include the likelihoods from other tests.

\item A set of decisions about the patient $0$ and their utilities
  conditional on the patient's health condition. We shall simply consider
  two decisions: dismiss ($\ynT$) or treat ($\yT$). See
  \sect~\ref{sec:decision_theory} for a summary of decision theory.
\end{itemize}

Our goal is to assign numerical values to the likelihoods for the health
conditions~\eqref{eq:test_prob_intro}: the conditional probability
distribution that the fMRI result of patient \enquote{$0$} is $\yf$ given
the health condition of that patient and all other data. In our notation,
\begin{equation}
    \label{eq:goal_1st}
    \begin{aligned}
  &\pf( \yF_0^{\yf} \| \yH_0^{\yh} \Land
  \yH_1^{\yh_1}\land \yF_1^{\yf_1}\Land
  \dotsb\Land
  \yH_{\yn}^{\yh_{\yn}}\land \yF_{\yn}^{\yf_{\yn}}\Land
    \yI )\,\di\yf
    \\
  \text{or just}\quad
    &\pf( \yF_0^{\yf} \| \yH_0^{\yh} \land \yD \land \yI)\,\di\yf.
    \end{aligned}
\end{equation}

A natural assumption helps us restrict the values the distribution above
may have. Within a group of patients \emph{having the same health
  condition} we assume that the probability that a patient shows a
particular fMRI $\yf_i$ does not depend on the particular value of the
patient's label $i$, no matter how many patients we have or may later add
in that health group. This assumption is called \emph{partial
  exchangeability}
\citep{definetti1938,diaconisetal1981,aldous1985,diaconis1988}. If the
labels carry \eg\ temporal or geographical information, partial
exchangeability means that we do not expect to observe particular kinds of
fMRI results more often in the future than in the past, or more frequently
in one location than another. As a concrete example: fix three possible
fMRI results $\yf_1$, $\yf_2$, $\yf_3$ (each is a vector with
$10^7$--$10^8$ positive components) and consider the fMRI tests of three
schizophrenic patients: say, one from five years ago in Germany, one from
last week in Scotland, and one to be done six months from now in Italy.
Partial exchangeability means that the probability that the German
patient's test gave $\yf_1$, the Scottish's gave $\yf_2$, and the Italian's
will give $\yf_3$, is numerically equal to the probability that the
German's gave $\yf_2$, the Scottish's $\yf_3$, and the Italian's will give
$\yf_1$; and likewise for all six possible permutations of the three
results. Keeping the same fixed fMRI results $\yf_1$, $\yf_2$, $\yf_3$, we
now consider three healthy patients instead, who may also live in different
times and places. Partial exchangeability means that also in this case the
values of the six possible joint probabilities obtained by permutation must
all be equal -- but this value can be different from the one for the
schizophrenic patients considered before. Hence the term \enquote{partial}: we
can freely exchange the joint results within the schizophrenic group and
within the healthy group without altering their probabilities, but not
across groups. This assumption extends in an analogous way to more
patients.

The assumption of partial exchangeability might not be completely true when
we consider geographical or epochal differences, but we may still consider
it as a good approximation. We are not making any exchangeability
assumptions about the probabilities of the health conditions of our
patients, though, because the incidence of a disease does often change with
time and can depend heavily on geographical location.

To express partial exchangeability mathematically, suppose that the
patients $i = 1,2,3,\dotsc$ have health condition $\yh=\yhu$ and the patients
$i' = 1',2',3',\dotsc$ health condition $\yh=\yhd$. Then the joint distribution for
their fMRI results satisfies 

\begin{multline}
  \label{eq:partial_exchangeability_example}
    \pf\Bigl( \tland_i\yF_{i}^{\yf_{i}}\ 
    \tland_{i'}\yF_{i'}^{\yf_{i'}}
    \bigcond
    \tland_i\yH_{i}^{\yhu}\ 
    \tland_{i'}\yH_{i'}^{\yhd} \Land \yI \Bigr) =
    \pf\Bigl( \tland_i\yF_{i}^{\yf_{\pi(i)}}\ 
    \tland_{i'}\yF_{i'}^{\yf_{\pi'(i')}}
    \bigcond
    \tland_i\yH_{i}^{\yhu}\ 
    \tland_{i'}\yH_{i'}^{\yhd} \Land \yI \Bigr)
  \\
  \text{for all permutations $\pi$ of $\set{i}\equiv\set{1,2,\dotsc}$, and all permutations
    $\pi'$ of $\set{i'}\equiv\set{1',2',\dotsc}$.}
\end{multline}

The assumption of partial exchangeability is simple and quite natural --
and very powerful: it implies, as shown by de~Finetti
\citetext{\citeyear{definetti1938}; \citealp[\sect~3]{diaconis1988};
  \citealp[\sect~4.6]{bernardoetal1994_r2000}}, that the joint
distributions above must have the form

\begin{equation}
  \label{eq:partial_exchangeability_theorem}
    \pf\Bigl(\tland_i \yF_{i}^{\yf_{i}}\ 
    \tland_{i'}\yF_{i'}^{\yf_{i'}}
    \bigcond
    \yH \land \yI \Bigr) =
    \iint \Bigl[ \prod_{i} \yph(\yf_{i})\Bigr]\,
    \Bigl[ \prod_{i'} \yps(\yf_{i'})\Bigr]\;
    \pf(\yph, \yps \|\yI)\;
    \di\yph\,\di\yps
\end{equation}
where $\yph$, $\yps$ are distributions over the possible values of $\yf$,
and $\pf(\yph,\yps \| \yI)$ is a \enquote{hyperdistribution} over such
distributions, determined by the assumptions $\yI$. The double integral
(which can be understood as a generalized Riemann integral:
\citealp{lamoreauxetal1998,swartz2001,kurtzetal2004}), is over all
distributions $\yph$, $\yps$. In other words, de~Finetti's theorem say that
the joint probability distribution for the fMRIs of healthy and
schizophrenic patients can be seen as the product of independent
distributions, identical for healthy cases and identical for schizophrenic
cases but different for the two cases, mixed over all possible such pairs
of distributions with weight $\pf(\yph,\yps \| \yI)$.

As a very cursory example, suppose we want the joint probability that a
healthy patient has fMRI result $\yf$ and a schizophrenic one $\yf'$.
De~Finetti's formula first tells us to consider all possible distributions
over positive vectors. As usual with infinities, \enquote{all} must be made
precise by specifying a topology \citep[for details see
\eg][]{definetti1938,diaconisetal1981,aldous1985,diaconis1988}; but
intuitively these distributions comprise, \eg, multivariate truncated
normals, gammas, exponentials, truncated Cauchys\ldots\ and innumerable
distributions that we can imagine and don't have a specific name for; all
with their possible parameter values. De~Finetti's formula tells us to
choose one distribution $\yph$, from all those possible ones, for the
healthy case and one $\yps$ for the schizophrenic case, and to attach a
weight to this pair, $\pf(\yph,\yps \| \yI)$; then to calculate this pair
at the values $\yf$, $\yf'$ and multiply them: $\yph(\yf)\times\yps(\yf')$.
Then we consider a new pair of distributions, attach a weight to them, and
again multiply their values at $\yf$ and $\yf'$. And so on, until all
possible pairs are considered. Finally we calculate the sum of all such
products, weighted accordingly:
$\int \yph(\yf)\yps(\yf')\,\pf(\yph,\yps \| \yI)\,\di\yf\,\di\yf'$.

The generalization of the formulae above to more than two health
conditions, or when only one health condition is considered, is
straightforward.

As the cursory example above made quite clear, an integral over probability
distributions is a mathematically complicated object
\citep[\cf][]{ferguson1974} and may not seem a great advancement in
assigning a value to the distribution~\eqref{eq:goal_1st}. In defence of
de~Finetti's formula we must say that it is completely manageable with
discrete data spaces and provides a great insight in the way we reason
about probability in relation to repeated events
\cite[\sect~4.2]{definetti1937,lindleyetal1976,kingman1978,kochetal1982,dawid2013,bernardoetal1994_r2000}.
It also has several important consequences for our inference, which we now
discuss.

Using de~Finetti's formula and the definition of conditional probability we
can rewrite our goal plausibility~\eqref{eq:goal_1st} as
\begin{subequations}\label{eq:goal_1st_exchangeability}
  \begin{gather}
    \pf( \yF_0^{\yf}\| \yH_0^{\yh} \land \yD \land \yI)
    = 
    \int \yp_{\yh}(\yf)\, \pf(\yph,\yps \|\yD \land \yI) \,\di\yph\,\di\yps
    \\[\jot]
    \text{with} \quad
    \pf(\yph,\yps \|\yD \land \yI) 
    = \frac{\bigl[ \prod_i \yp_{\yh_i}(\yf_i) \bigr]\,
      \pf(\yph,\yps \|\yI)
    }{
      \int \bigl[ \prod_i \yp_{\yh_i}(\yf_i) \bigr]\,
      \pf(\yph,\yps \|\yI) \,\di\yph\,\di\yps }    
      .
    \label{eq:updated_hyperprior_nonparametric}
  \end{gather}
\end{subequations}

The latter is called posterior distribution since it is conditional on all
data $\yD$.

Excluding pathological prior distributions \citep{diaconisetal1986}, this
posterior distribution becomes more and more concentrated on two particular
distributions $\bigl(\ypc_{\yhu}, \ypc_{\yhd}\bigr)$, fully determined by
the data, as our data $\yD$ comprise a larger and larger number of
patients. This concentration occurs independently of the original shape of
the distribution $\pf(\yph,\yps \|\yI)$. In this limit our probability
distribution~\eqref{eq:goal_1st} becomes
\begin{equation}
  \label{eq:goal_1st_limit}
  \pf( \yF_0^{\yf} \| \yH_0^{\yh} \land \yD \land \yI)
  \approx
  \ypc_{\yh}(\yf)
\end{equation}
with $\ypc_{\yh}$ completely determined by the data $\yD$. This would solve
our plausibility assessment \eqref{eq:goal_1st}.

De~Finetti's formula therefore tells us also the theoretical limit by which
the pre-test probability for the health condition of the patient,
$\p(\yH_0^{\yh} \| \yD \land \yI)$, can be improved by the fMRI result. For
example, if $\ypc_{\yh}(\yf)$ is more or less uniform in $\yf$ or has the
same peaks in $\yf$ for each $\yh$, then the fMRI is of no use for
discriminating the health condition of the patient. This result follows
mathematically from the assumption of
exchangeability~\eqref{eq:partial_exchangeability_example} and the rules of
probability calculus, hence no amount of ingenuity could overcome this
limit.

In our case the amount of data $\yD$ is not enough to allow the use of the
approximation~\eqref{eq:goal_1st_limit}. We should use the general
formula~\eqref{eq:goal_1st_exchangeability}, but it is unwieldy in two
respects. First, the fMRI result $\yf$ of a patient is a positive-valued
vector with $10^7$--$10^8$ components or more \citep{lindquist2008}, so the
distributions $\yph$, $\yps$ are highly multidimensional. Second, the
formula asks us to consider in principle \emph{all} such distributions, as
explained in the example above.

We tame this double unwieldiness in two ways. First, it is conceivable that
not all information contained in the fMRI result $\yf$ of a patient be
relevant to discriminate the patient's health condition $\yh$. The integral
in \eqn~\eqref{eq:goal_1st_exchangeability}, if it could be performed,
would automatically winnow out the relevant information
\citep[\chap~17]{jaynes1994_r2003}, possibly reducing the problem to a
lower-dimensional set in the space of fMRI data. Being unable to perform
the integral, we must try to apply heuristics based on our understanding of
the target conditions and perform such dimensional reduction by hand. For
example, by employing the hypothesis that in schizophrenic patients the
time correlation between brain regions is altered with respect to healthy
ones. We can thus address the first problem by reducing fMRI data $\yf$ to
a manageable set of graph properties $\yx$, and applying our inference
directly on these, as explained in the next section.

Second, we entertain a working assumption about which features of our graph
data $\set{\yx_i}$ from a set of patients are relevant for inferences about
new patients. We can, for example, assume that only the first and second
moments are relevant for making predictions about the graph quantities to
be observed in the new patients; these moments are then called
\emph{sufficient statistics}. Assumptions of this kind reduce the
infinite-dimensional space all possible distributions $(\yph,\yps)$ to a
finite-dimensional space of a parametric family of distributions of
exponential type, as explained in \sect~\ref{sec:sufficiency}.

\subsection{Trimming the data space: functional connectivity}
\label{sec:data_reduction}

The preprocessed functional image in standard space, in which the activity for each voxel is recorded,  consists of approximately \(10^{17}\)  time series of $140$ time points.  We reduce this huge data space by the following steps. 

First, we consider only the activity of the voxels belonging to the $94$
regions defined by the lateral cortical Oxford atlas (see
\sect~\ref{sec:data_acquisition} and \citealp{Desikan2006}) and average the
activity of all voxels in a region (details in
\sect~\ref{sec:ROI_selelection_and_conn_measure}). Second, we measure the
functional connectivity defined by the Pearson correlation coefficient
between pairs of regions, obtaining \(94\times 93/2=4371\) connectivity
weights. This is still a considerable data space for our computational
resources, so we select $\yd=40$ connectivity weights that exhibit the
greatest difference in their connection weight average across the
schizophrenic and healthy groups.

\begin{figure}[!h]
  \centering
\includegraphics[height=0.37\linewidth]{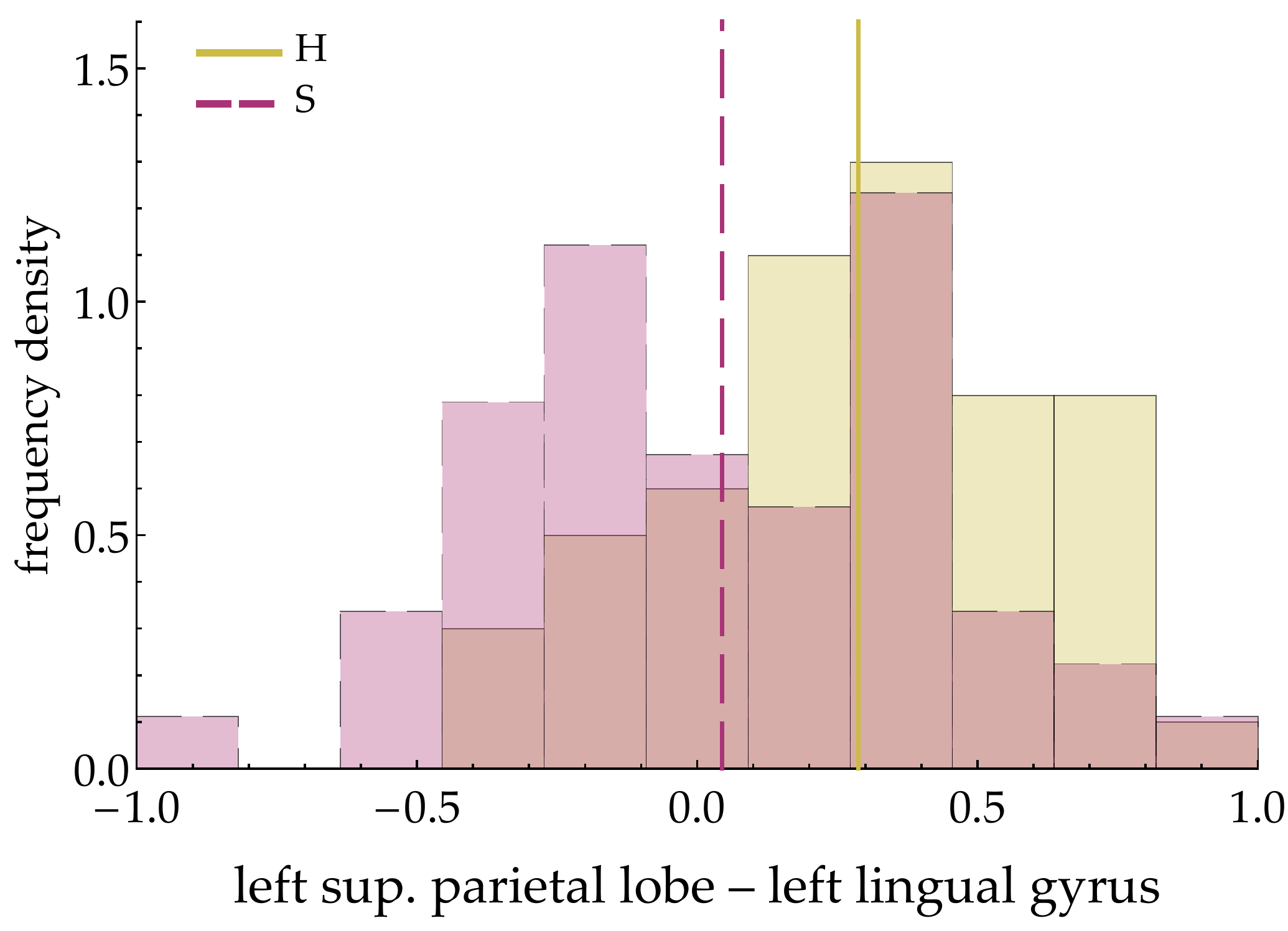}\hfill
\includegraphics[height=0.37\linewidth]{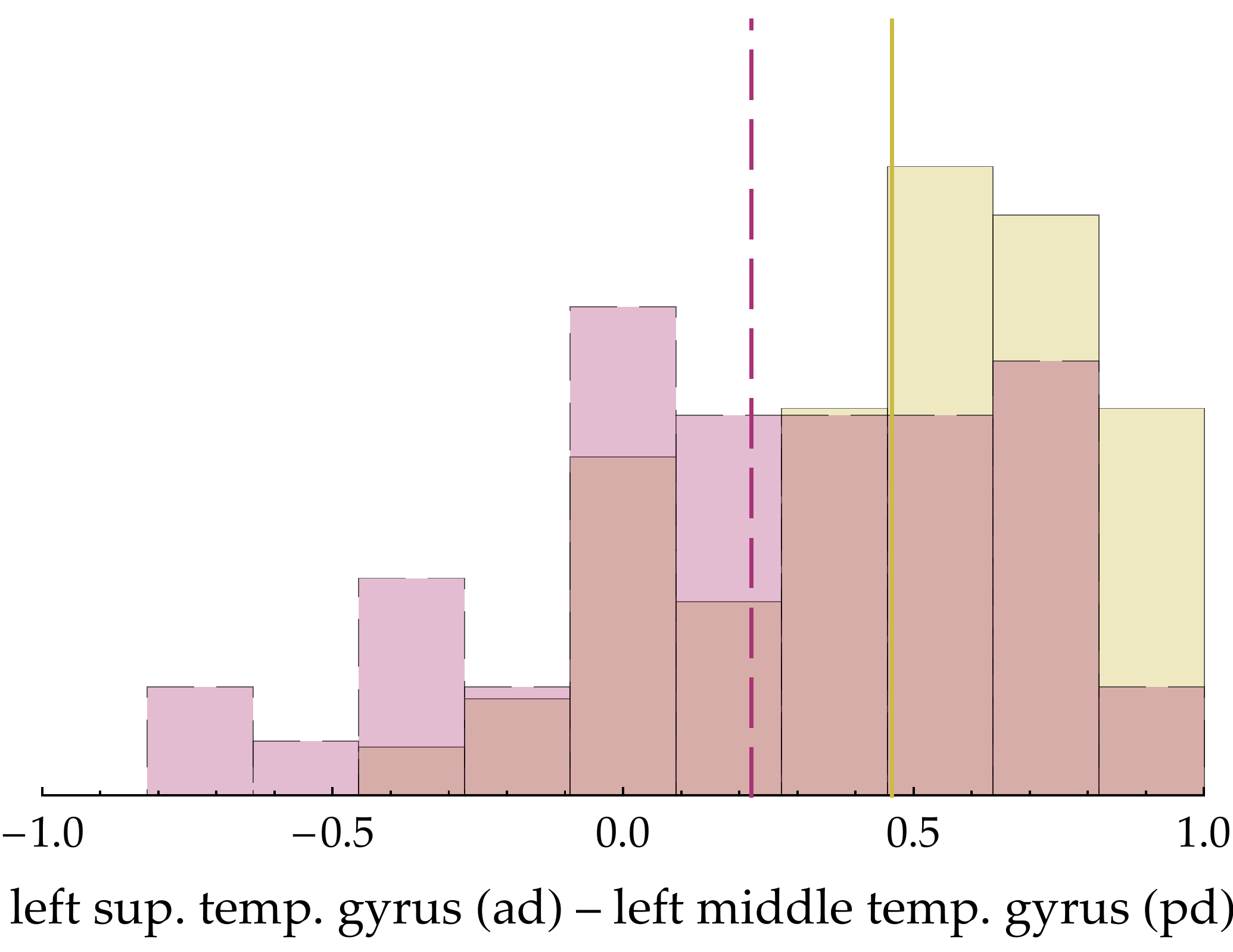}\\[1\jot]
\includegraphics[height=0.37\linewidth]{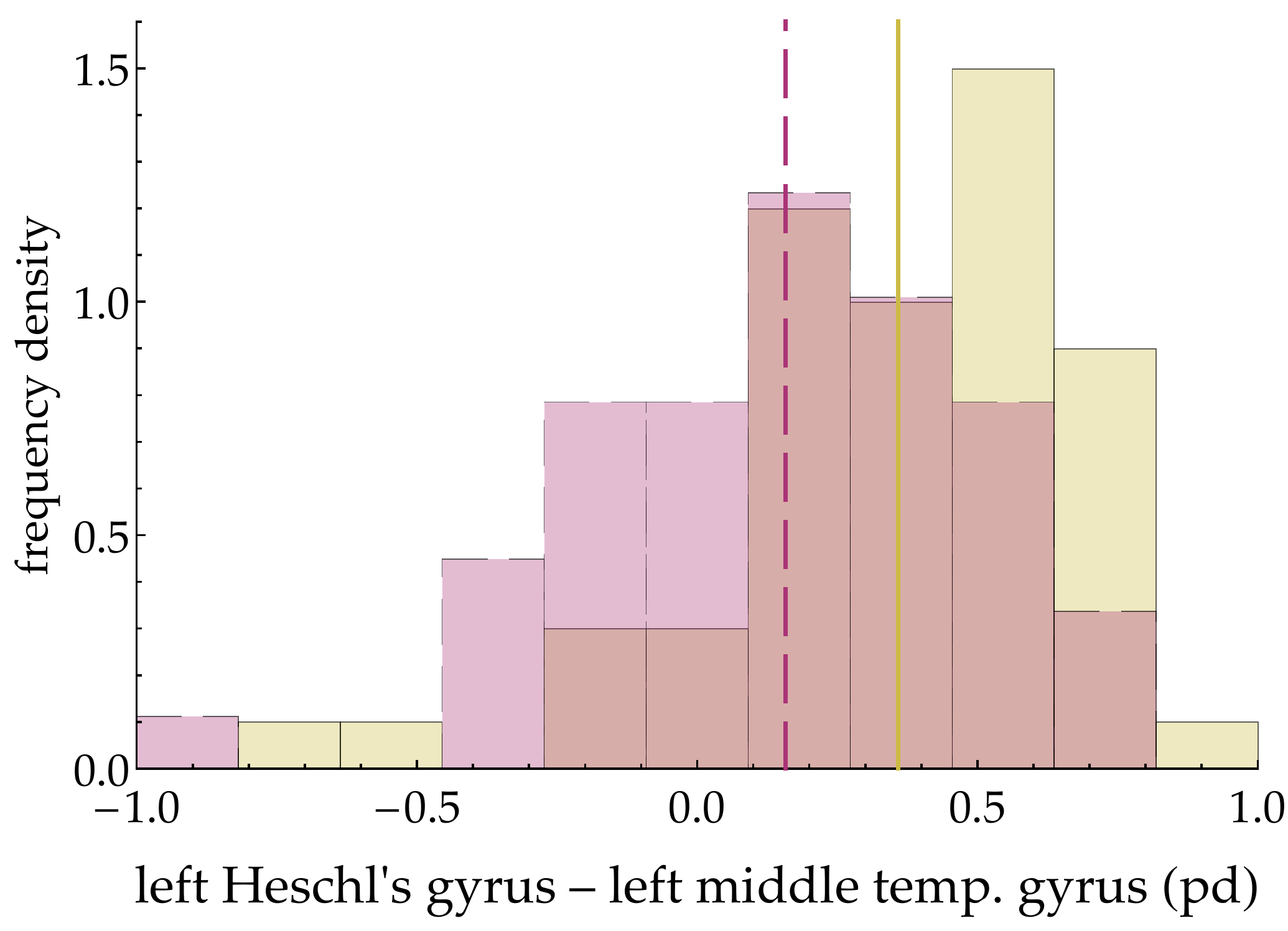}\hfill
\includegraphics[height=0.37\linewidth]{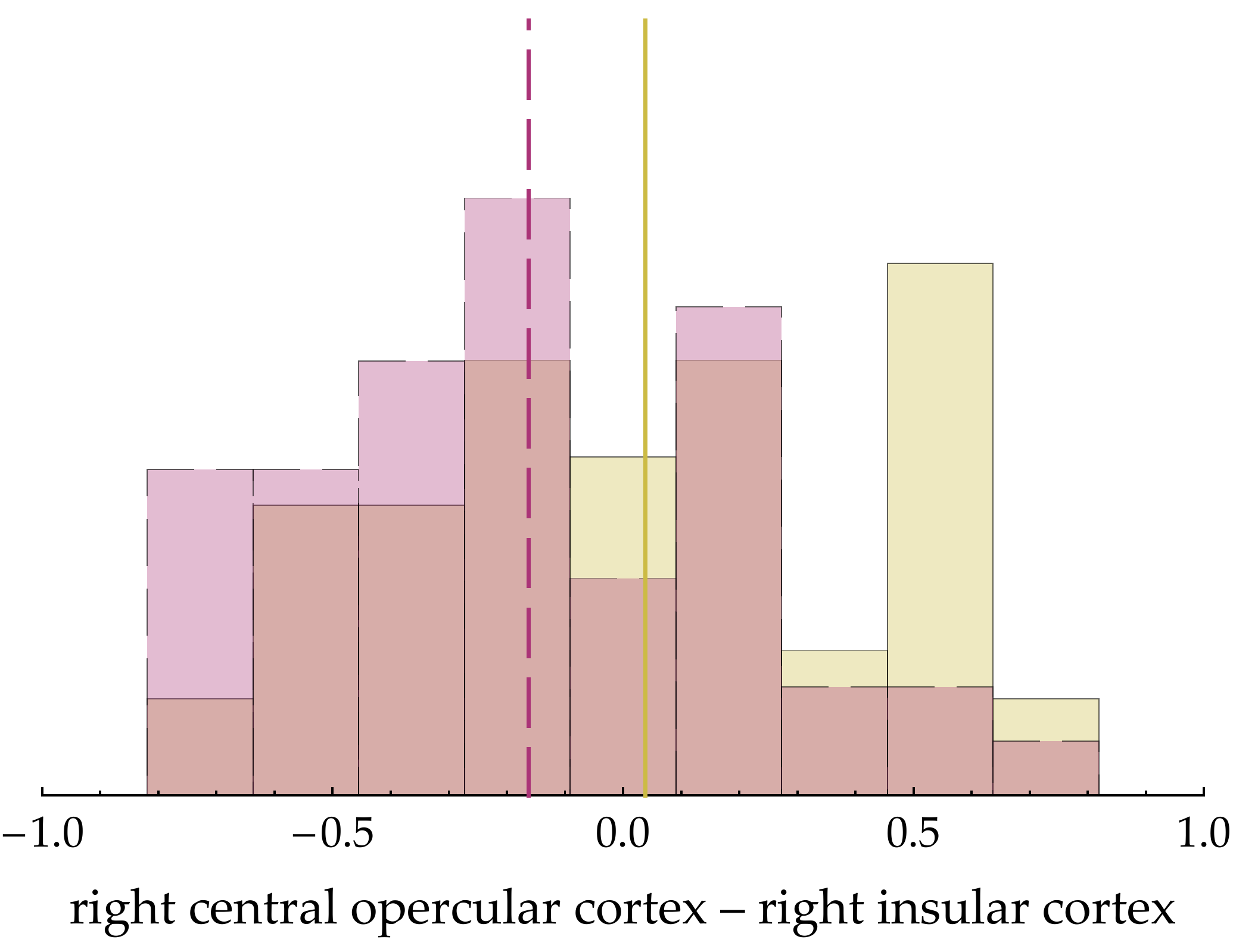}\\
\caption{\textbf{Distributions of functional connectivities of
    schizophrenic and healthy patients.} Normalized density histograms of
  connectivity weights for healthy ($\yhu$, yellow, solid) and
  schizophrenic ($\yhd$, red, dashed) patients of four cortical connections
  selected to demonstrate our statistical framework. Empirical means are
  shown as vertical lines. All connectivities for healthy and schizophrenic
  patients have substantial overlap, evidenced by the darker regions in the
  histograms.
}
\label{fig:overlapping_distribution}
\end{figure}

The resulting distributions for four of these connectivity weights are
depicted in \fig~\ref{fig:overlapping_distribution}. Note that for each
considered brain connection, the histograms of the healthy group and the
schizophrenic group display significant overlap, such that none of them
could be used in isolation to reliably discriminate between two groups.

Our data space is therefore vastly reduced, from  $\clop{0,\infty}^{10^{17}}$

to $\clcl{-1,1}^{40}$. We let the symbol $\yf$ stand for the set of
connectivity weights extracted from the fMRI data, rather than the full
fMRI data themselves. With this new meaning of the symbol $\yx$, the
likelihoods for the health conditions~\eqref{eq:goal_1st} and de~Finetti's
formulae~\eqref{eq:partial_exchangeability_theorem}--\eqref{eq:goal_1st_exchangeability}
remain formally unchanged, but now involve a data space with much fewer
dimensions.

\subsection{Trimming the distribution space: models by sufficiency and
  generalized normals}
\label{sec:sufficiency}

\subsubsection{Parametric models}
\label{sec:parametric_models}

The integrals in de~Finetti's
formulae~\eqref{eq:partial_exchangeability_theorem}--\eqref{eq:goal_1st_exchangeability}
still represent a mixture of all imaginable pairs of probability
distributions $(\yph, \yps)$ over the space $\clcl{-1,1}^{\yd}$, and are
therefore extremely complex. We now examine two ways to reduce this
integral to a manageable set of distributions and to obtain analytically
tractable formulae.

In the Bayesian literature, the complication of considering all possible
distributions $\yp_{\yh}(\yx)$ of the quantity $\yx$ is typically
sidestepped by restricting them to a finite-dimensional set of
distributions $\yL_{\yh}(\yx \| \yth_{\yh})$, identified or indexed by a
finite number of parameters $\yth_{\yh}$. For this reason, such a set is
called a parametric family of distributions. An example of parametric
family is the set of $\yd$-variate normal distributions parameterized by
their mean $\bm{\mu}$ and covariance matrix $\bm{\varSigma}$. With this
restriction, the integrals in de~Finetti's
formulae~\eqref{eq:partial_exchangeability_theorem} and
\eqref{eq:goal_1st_exchangeability} represent mixtures of distributions
within the parametric family, the weight for each distribution being
represented by a weight for its parameters. That is, we are no longer
considering mixtures of all possible multivariate truncated normals,
gammas, exponentials, \etc, as in the example of
\sect~\ref{sec:exchangeability}, but only mixtures of truncated normals,
say, with different means and covariance matrices. These integrals are thus
ordinary finite-dimensional integrals. What happens to
formulae~\eqref{eq:partial_exchangeability_theorem}--\eqref{eq:goal_1st_exchangeability}
is that
\begin{equation}
  \label{eq:replacement_by_model}
  \yp_{\yh}(\yx) \text{ is replaced by } \yL_{\yh}(\yx \|\yth_{\yh}), \qquad
  \pf(\yph,\yps \| \yI)\,\di\yph\,\di\yps \text{ is replaced by }
  \pf(\ythh,\yths \| \yM,\yI)\,\di\ythh\,\di\yths.
\end{equation}
The distribution $\yL_{\yh}$ is called the likelihood of the parameters
$\yth_{\yh}$ , and $\pf(\ythh,\yths \| \yM, \yI)$ is the prior parameter
distribution. A parametric family and a prior distribution over its parameters
are jointly called a parametric statistical model, which we denote by
$\yM$. The term \enquote{model} is justly criticized by some probability
theorists \citep[see \eg\ Besag \amp\ Kalman in][]{besagetal2002} but widely
used, so we shall adopt it here.

In our present problem, the parametric statistical model needs not be the
same for all health conditions: for example, we could use normal
distributions for one condition and beta distributions for another, if that choice better reflected the distributions of connectivity weights under the two
different health conditions. For this reason, we use the subscript
\enquote{$\yh$} in the formulae above. The likelihood we want to determine,
\eqn~\eqref{eq:goal_1st},
thus becomes, from \eqn~\eqref{eq:goal_1st_exchangeability} with the
replacements~\eqref{eq:replacement_by_model},
\begin{subequations}\label{eq:goal_1st_exchangeability_parametric}
  \begin{gather}
    \pf( \yF_0^{\yf}\| \yH_0^{\yh} \land \yD \land \yM \land \yI)
    = 
    \int \yL_{\yh}(\yf \| \yth_{\yh})\,
    \pf(\ythh,\yths \|\yD , \yM ,\yI) \,\di\ythh\,\di\yths
    \\[\jot]
    \text{with} \quad
    \pf(\ythh,\yths \|\yD ,\yM , \yI) 
    = \frac{\bigl[ \prod_i \yL_{\yh_i}(\yf_i \|\yth_{\yh_i}) \bigr]\,
      \pf(\ythh,\yths \|\yM , \yI)
    }{
      \int \bigl[ \prod_i \yL_{\yh_i}(\yf_i \| \yth_{\yh_i}) \bigr]\,
      \pf(\ythh, \yths \|\yM ,\yI) \,\di\ythh\,\di\yths}    
    .
    \label{eq:updated_hyperprior}
  \end{gather}
\end{subequations}

\subsubsection{Models by sufficient statistics}
\label{sec:suff_stat}

The choice of a statistical model often appears as an art and a matter of
experience. Notable statisticians have voiced concerns over the esoteric
character of this choice. Dawid~\citey[p.~220]{dawid1982} says:
\enquote{Where do probability models come from? To judge by the resounding
  silence over this question on the part of most statisticians, it seems
  highly embarrassing}. And Diaconis \citey[\sect~8, p.~121]{diaconis1988} remarks:
\begin{quotation}
  de Finetti's alarm at statisticians introducing reams of unobservable
  parameters has been repeatedly justified in the modern curve fitting
  exercises of today's big models. These seem to lose all contact with
  scientific reality focusing attention on details of large programs and
  fitting instead of observation and understanding of basic mechanism. It
  is to be hoped that a fresh implementation of de Finetti's program based
  on observables will lead us out of this mess.
\end{quotation}
Authors like these have also tried to develop intuitive methods to choose a
model, for example by proving that a parametric family can be uniquely
determined by some symmetry assumptions about our inferences, or by other
information-theoretical properties
\citetext{\citealp[\chap~4]{bernardoetal1994_r2000};
  \citealp[\sect~5.5]{lindley1965b_r2008}; an enlightening discussion of
  this topic is given by \citealp{dawid2013}}. In the present study we want
to emphasize, as Cifarelli \amp\ Regazzini \citey{cifarellietal1982} did,
that a statistical model can be chosen by selecting a \emph{sufficient
  statistics} \citep[and the textbook references
above]{kolmogorov1942,freedman1962b,diaconisetal1980b,diaconisetal1981,cifarellietal1982,lauritzen1982_r1988,diaconis1992,kallenberg2005}.
Here is an example.

Imagine that we have patients labelled $i'\in\set{1',2',\dotsc}$, and $\yn$
patients labelled $i\in\set{1,2,\dotsc}$, all with the same health
condition. Of the second set of patients we also know the connectivity
weights $\set{\yx_i}$ obtained by fMRI. We want to specify the joint
probability distribution $\pf( \set{\yx_{i'}}\| \set{\yx_i}, \yI)$ that the
fMRIs of the patients $\set{i'}$ yield connectivity weights
$\set{\yx_{i'}}$, conditional on our knowledge of the connectivity weights
of the $\yn$ patients $\set{i}$. Now assume that the probabilities for the
fMRI results are exchangeable, so that de~Finetti's
formulae~\eqref{eq:partial_exchangeability_theorem}
and~\eqref{eq:goal_1st_exchangeability} hold. Also assume that in order to
specify the joint distribution we do not need the full set of data
$\set{\yx_i}$, but only their number $\yn$ and some statistics, \eg\ the
empirical mean and covariance matrix of these data,
\begin{equation}
  \label{eq:example_suff_stat}
  \av{\yx} \defd\frac{1}{\yn}\tsum_i \yx_i, \qquad
  \Cov(\yx) \defd \frac{1}{\yn}\tsum_i(\yx_i-\av{\yx})(\yx_i-\av{\yx})\T;
\end{equation}
the rest of the details of the data $\set{\yx_i}$ being irrelevant. In
other words, we are assuming that the statistics above are
\emph{sufficient} for us to make predictions as if we had the full data. In
symbols,
\begin{equation}
  \label{eq:sufficiency_example}
  \pf( \set{\yx_{i'}}\|  \set{\yx_i}, \yI)
  = 
  \pf(\set{\yx_{i'}} \|  \yn, \av{\yx}, \Cov{\yx}, \yI).
\end{equation}
If this is true no matter the number of patients $\set{i'}$ and $\set{i}$,
then these statistics are called \emph{(predictive) sufficient statistics}.
There are several notions of sufficiency, including the traditional one by
Fisher \citey{fisher1922} and Neyman \citey{neyman1935}, but they all are
more or less equivalent \citep[\sect~4.5.2]{bernardoetal1994_r2000}.

The assumption of the existence of sufficient statistics has a very
important consequence for de~Finetti's
formulae~\eqref{eq:partial_exchangeability_theorem}
and~\eqref{eq:goal_1st_exchangeability}: the space of possible prior
distributions is hugely reduced, constrained to be non-zero only over a
parametric family of distributions that is determined by the sufficient
statistics. The replacement~\eqref{eq:replacement_by_model} takes place,
leading to the simpler
formula~\eqref{eq:goal_1st_exchangeability_parametric} for the likelihoods
of the health conditions. The number of parameters is equal to that of the
sufficient statistics. The proof of this reduction was given by Pitman and
Koopman \citetext{\citealp{koopman1936,pitman1936,darmois1935}; for
  generalizations, \eg\ to the discrete case, see
  \citealp{hipp1974,andersen1970,denny1967,fraser1963,barankinetal1963}}.
When the sufficient statistics are the mean and covariance matrix, as
above, the likelihoods turn out to be (truncated) multivariate normal
distributions.

In the slightly more complicated case of two or more health conditions and
the assumption of partial
exchangeability~\eqref{eq:partial_exchangeability_example}, this theorem
leads to formula~\eqref{eq:goal_1st_exchangeability_parametric} with
likelihoods $\yL_{\yh}$ determined by the sufficient statistics that we
have chosen for the different health conditions
\citep[\sect~4.6]{bernardoetal1994_r2000}. The prior distribution over the
parameters of the likelihoods, $\pf(\ythh,\yths \| \yI)$, is not determined
by the theorem, but has to be determined by other consideration that can
again involve symmetry and information theory.

As we mentioned in the introduction, the notion of sufficient statistics
can be a helpful bridge between biophysical considerations and the
specification of probabilities. It may be easier to conceive and understand
a connection between biophysical considerations and some statistics of our
measurements, than between biophysical considerations and an abstract
multidimensional distribution function. Once such statistics are selected,
they in turn uniquely select a probability distribution for us. Vice versa,
if a statistical model based on some sufficient statistics proves to be
very reliable in its predictions, we may conclude that its sufficient
statistics must have an important biological meaning.

In the rest of this study we shall use three statistical models determined by
three different sufficient statistics. Our choice of statistics is
unfortunately not biologically motivated, as such models have yet to be determined for fMRI data. However, they are adequate to demonstrate the approach and we hope that authors with more experience will pursue this
line of thought and derive better-motivated sufficient statistics.

\subsubsection{Edgeworth's \enquote{method of translation}: generalized normal
  models}
\label{sec:generalized_normals}

The assumption of a sufficient statistics makes the integrals in
de~Finetti's formulae~\eqref{eq:partial_exchangeability_theorem}
and~\eqref{eq:goal_1st_exchangeability} finite-dimensional, but these
integrals and other expressions that depend on them, like the post-test
probabilities, may still lack a closed form and be analytically
intractable. In this case we could use numerical methods, a computationally
costly possibility we consider in the Discussion,
\sect~\ref{sec:improvements}. In the present work we choose models with
closed-form formulae instead; their swift calculation facilitates the model
comparison to be illustrated later.

Our starting point is an analytically tractable model by sufficient
statistics that has been the subject of much study
\citetext{\citealp[\sect~3.6]{gelmanetal1995_r2014};
  \citealp{minka1998_r2001,murphy2007}}: it has a normal likelihood, with
mean $\ylmm$ and covariance matrix $\ylss$ as parameters, and a
normal-inverse-Wishart prior distribution over these parameters. This
parameter prior maintains the same functional form when updated with
training data; this kind of prior is called \emph{conjugate}
\citetext{\citealp[\chap~9]{degroot1970_r2004};
  \citealp{diaconisetal1979b}}. This model is outlined more in detail in
\sect~\ref{sec:normal-inverse-wishart}.

We try to make the normal~+ normal-inverse-Wishart model more flexible by
combining it with an idea that Edgeworth \citey{edgeworth1898} called
\enquote{Method of Translation}, discussed also by Johnson
\citey{johnson1949} and Mead \citey{mead1965}: the transformation of a
quantity into a normally distributed one. That is, instead of considering
the connectivity weights $\yx$, we consider transformed quantities
$\ytr(\yx)$, where $\ytr$ is a component-wise monotonic function, and
suppose the latter quantities to be normally distributed. This leads to
generalized-normal likelihoods of the form
\begin{equation}\label{eq:generalized_normal}
\dnormal[\ytr(\yx) \| \ylmm, \ylss] \,\ytr'(\yx) \,\di\yx
\end{equation}
where $\dnormal$ is the normal distribution with mean $\ylmm$ and covariance
matrix $\ylss$, and $\ytr'$ is the Jacobian determinant of $\ytr$.

This simple idea has an amazing scope: it allows us to explore a vast range
of non-normal likelihoods -- in particular likelihoods defined on bounded
domains such as $\clcl{-1,1}^{\yd}$ -- and to keep the low computational
costs of the conjugate prior. In our present problem it has also an
additional convenient feature: in the calculation of the post-test
probability, the Jacobian determinant $\ytr'$ disappears, as
\eqn~\eqref{eq:post-test_final} below shows.

These generalized-normal models, which we denote by $\yMl$, are also
determined by a choice of sufficient statistics, analogous to
\eqn~\eqref{eq:example_suff_stat}: the mean and covariance matrix of the
transformed data $\set{\ytr(\yx_i)}$,
\begin{equation}
  \label{eq:suff_stat_gen_normal}
  \av{\ytr(\yx)}, 
  \qquad
  \Cov[\ytr(\yx)]. 
\end{equation}

\bigskip

In our partially exchangeable case, with
formulae~\eqref{eq:goal_1st_exchangeability_parametric}, we need to specify
a likelihood $\yL_{\yh}$ for each health condition $\yh$. We assume these
two likelihoods $\yL_{\yhu}$, $\yL_{\yhd}$ to be functionally identical
generalized normals, \ie\ the function $\ytr$ is the same for the healthy
and the schizophrenic case; but their means and covariance matrices
$(\ylmm_{\yhu},\ylss_{\yhu})$ and $(\ylmm_{\yhd},\ylss_{\yhd})$ can be
different.

We also need to specify a joint parameter prior for
$(\ylmm_{\yhu},\ylss_{\yhu}; \ylmm_{\yhd},\ylss_{\yhd})$. To use the
analytic advantage of the conjugate prior, we assume that the distribution
for these parameters is a product of independent distributions:
\begin{equation}
  \label{eq:logit-normal_parameters_factor_condition}
  \pf(\ylmm_{\yhu},\ylss_{\yhu} ;
  \ylmm_{\yhd},\ylss_{\yhd}\| \yMl, \yI)
  = \pf(\ylmm_{\yhu},\ylss_{\yhu} \| \yMl, \yI)\times
  \pf(\ylmm_{\yhd},\ylss_{\yhd}  \| \yMl, \yI),
\end{equation}
each of them being a normal-inverse-Wishart distribution described in
\sect~\ref{sec:normal-inverse-wishart}. This independence assumption is
quite strong and has an important consequence: \emph{the likelihood for a
  health condition only depends on the data from previous patients having
  that same health condition}.

With the assumptions above, the likelihood for the health condition needed
by the clinician has a closed form for this model
(see \sect~\ref{sec:normal-inverse-wishart}). The likelihood for patient $0$'s
being healthy, in view of the patient's measured connectivity weights $\yx$,
and given the data $(\yx_i)$ from previous $\ynh$ healthy patients, is
\begin{shaded}
  \begin{equation}
    \label{eq:final_likelihood}
    \pf(\text{fMRI result $\yx$} \Cond \text{healthy} \Land
    \text{prior info})
    =
    \dstudentt\Bigl[\ytr(\yx) \bigcond \ynu_{\yhu}-\yd+1, \ymu_{\yhu},
    \tfrac{\yka_{\yhu}+1}{\yka_{\yhu}\,(\ynu_{\yhu}-\yd+1)}\yLa_{\yhu} \Bigr]
    \,\oldprod_k\ytr'(\yxx_k),
  \end{equation}
\end{shaded}
\noindent where $\dstudentt$ is a multivariate t~distribution with
$\ynu_{\yhu}-\yd+1$ degrees of freedom, mean $\ymu_{\yhu}$, and scale
matrix $\frac{\yka_{\yhu}+1}{\yka_{\yhu}\,(\ynu_{\yhu}-\yd+1)}\yLa_{\yhu}$.
This distribution has covariance matrix
$\frac{\yka_{\yhu}+1}{\yka_{\yhu}\,(\ynu_{\yhu}-\yd-1)}\yLa_{\yhu}$ (note
the different denominator from the scale matrix), and approaches a
generalized normal for large $\ynu_{\yhu}$. The final factor is the Jacobian
determinant of $\ytr$.

The most important feature of this likelihood is the dependence of the
coefficients $(\yka_{\yhu}, \ymu_{\yhu}, \ynu_{\yhu}, \yLa_{\yhu})$ on the
data $(\yx_i)$ of the previous $\ynh$ healthy patients:
\begin{equation}
  \label{eq:update_coeffs_healthy}
    \begin{aligned}
    \yka_{\yhu} &= \ykao + \ynh,\quad     &\ynu_{\yhu} &= \ynuo + \ynh,\\
    \ymu_{\yhu} &= \frac{\ykao\,\ymuo+\ynh\, \av{\ytr(\yx)}}{\ykao+\ynh},\quad
    &\yLa_{\yhu} &= \yLao + \ynh\, \Cov[\ytr(\yx)]
           + \frac{\ykao\,\ynh}{\ykao+\ynh} \bigl[\av{\ytr(\yx)}-\ymuo\bigr] \bigl[\av{\ytr(\yx)}-\ymuo\bigr]\T,
  \end{aligned}
\end{equation}
where $(\ykao,\ymuo,\ynuo,\yLao)$ are prior coefficients that represent the
clinician's knowledge before any patients were observed. As the number
$\ynh$ of observed healthy patients increases, the probability for the
transformed data $\ytr(\yx)$ tends to a normal distribution with mean and
covariance matrix equal to the empirical average and covariance matrix of
the transformed data. The formulae above show that previous data enter only
through the sufficient statistics $\av{\ytr(\yx)}$ and $\Cov[\ytr(\yx)]$.

An analogous formula holds for the likelihood for the patient's being
schizophrenic, with coefficients
$(\yka_{\yhd}, \ymu_{\yhd}, \ynu_{\yhd}, \yLa_{\yhd})$ that depend on the
data of previous schizophrenic patients and some initial coefficients. The
function $\ytr$ and the prior coefficients $(\ykao,\ymuo,\ynuo,\yLao)$
could be different for the healthy and schizophrenic cases, but for
simplicity we assume them identical for both health conditions.

\bigskip

If $(\ypph,\ypps)$ is the pre-test probability distribution for the health
condition of patient $0$, his post-test probability to be healthy is
\begin{shaded}
  \begin{multline}
    \label{eq:post-test_final}
    \p(\text{healthy}\Cond \text{fMRI result}\Land\text{prior info})
    ={}\\[2\jot]
    \frac{\dstudentt\Bigl[\ytr(\yx) \bigcond \ynu_{\yhu}-\yd+1, \ymu_{\yhu},
      \tfrac{\yka_{\yhu}+1}{\yka_{\yhu}\,(\ynu_{\yhu}-\yd+1)}\yLa_{\yhu} \Bigr]
      \,\ypph}{
      \dstudentt\Bigl[\ytr(\yx) \bigcond \ynu_{\yhu}-\yd+1, \ymu_{\yhu},
      \tfrac{\yka_{\yhu}+1}{\yka_{\yhu}\,(\ynu_{\yhu}-\yd+1)}\yLa_{\yhu} \Bigr]
      \,\ypph
      +
      \dstudentt\Bigl[\ytr(\yx) \bigcond \ynu_{\yhd}-\yd+1, \ymu_{\yhd},
      \tfrac{\yka_{\yhd}+1}{\yka_{\yhd}\,(\ynu_{\yhd}-\yd+1)}\yLa_{\yhd} \Bigr]
      \,\ypps
    }.
  \end{multline}
\end{shaded}
\noindent Note that the Jacobian determinants $\ytr'$ do not appear in this
formula.

\subsubsection{Generalized normal models in our study}
\label{sec:our_models}

In the rest of our study we compare three different transformations $\ytr$
of the connectivity weights $\yx$, with one set of prior coefficients
$(\ykao,\ymuo,\ynuo,\yLao)$ each:
\begin{description}[wide]
\item[Logit-normal model: ] A slightly modified logit transformation
  \begin{equation}
  \label{eq:logit_tr}
\ytr(\yxx_k) \defd \ln\frac{1+\yxx_k}{1-\yxx_k},
\qquad
\ytr'(\yxx_k) = \frac{2}{1-\yxx_k^2},
\end{equation}
with prior coefficients
  \begin{equation}
    \label{eq:logit_tr_consts_flat}
    \ykao = 1,\quad \ymuo=0,\quad\ynuo=\yd+1,\quad\yLao = (\yd+2)\id_{\yd}.
  \end{equation}

\item[Tangent-normal model: ] A tangent transformation
  \begin{equation}
  \label{eq:tan_tr}
  \ytr(\yxx_k) \defd \tan\frac{\pu \yxx_k}{2},
\qquad
\ytr'(\yxx_k) = \frac{\pu}{1+\cos\pu\yxx_k},
\end{equation}
with prior coefficients
  \begin{equation}
    \label{eq:tan_tr_consts_flat}  
    \ykao = 1,\quad \ymuo=0,\quad\ynuo=\yd+1,\quad\yLao = \frac{\yd+2}{4}\id_{\yd}.   
  \end{equation}

\item[Normal model: ] An identity transformation (that is, no transformation at all)
  \begin{equation}
  \label{eq:id_tr}
  \ytr(\yxx_k) \defd \yxx_k,
\qquad
\ytr'(\yxx_k) = 1,
\end{equation}
with prior coefficients
  \begin{equation}
    \label{eq:id_tr_consts_flat}  
    \ykao = 1,\quad \ymuo=0,\quad\ynuo=\yd+1,\quad\yLao = 10\id_{\yd}. 
  \end{equation}
\end{description}
For brevity we shall denote
$\ytr(\yx) \defd \bigl( \ytr(\yxx_i) \bigr)$.

\begin{figure}[!t]
  \centering
\includegraphics[width=0.75\linewidth]{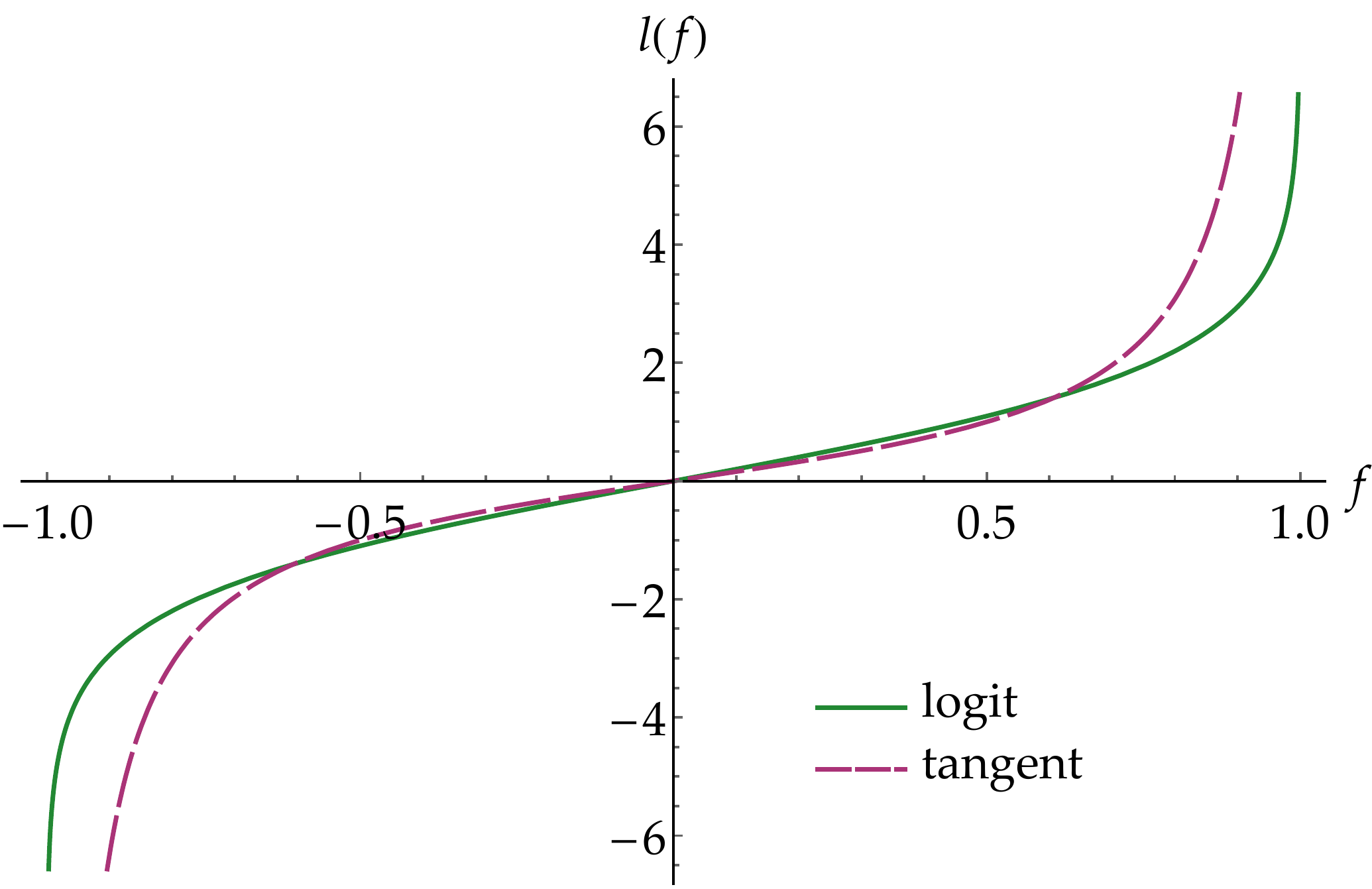}\\[10\jot]
\includegraphics[width=0.75\linewidth]{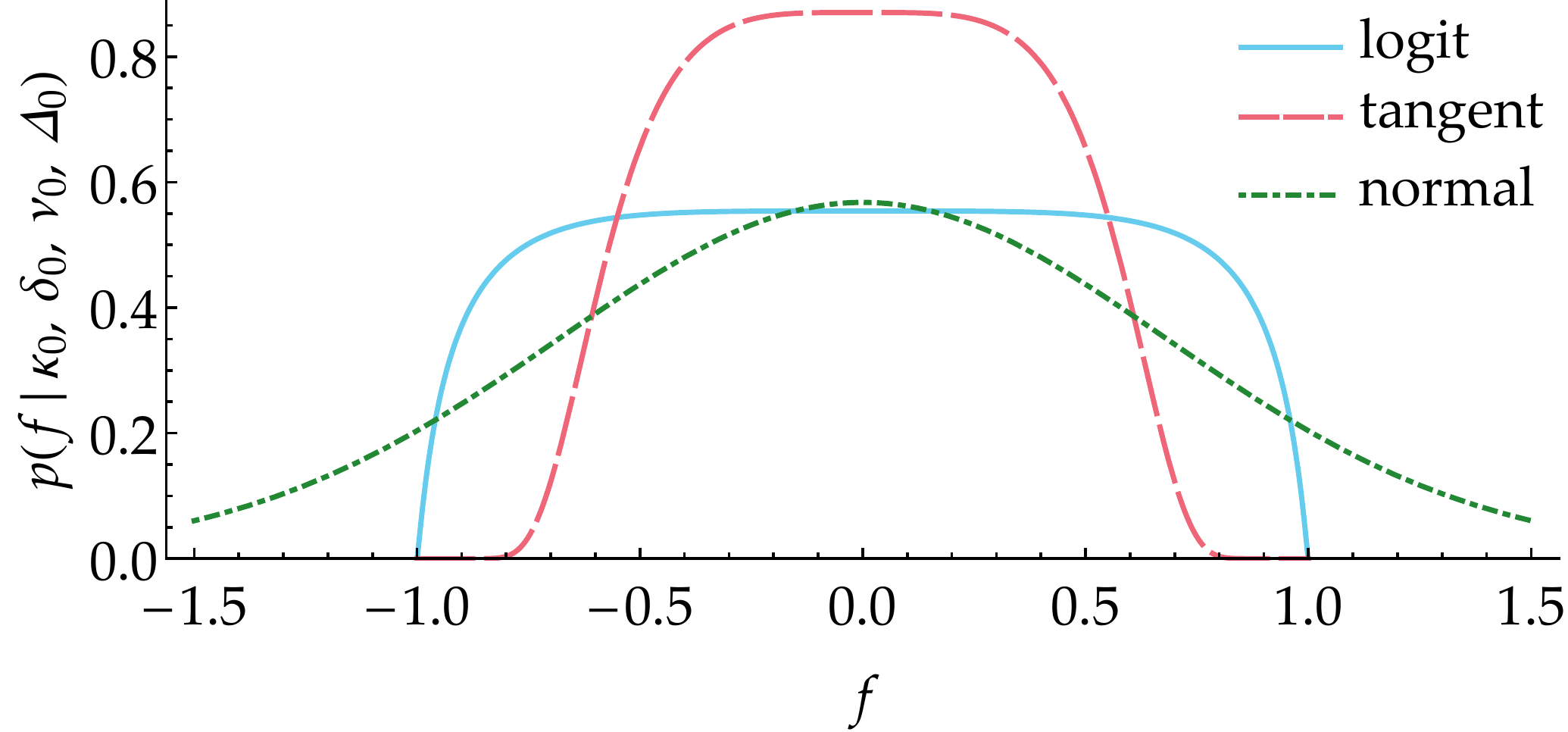}
\caption{General normalized models. Upper panel: Generalized logit and tangent transformation
  functions as defined in the text. Lower panel: Prior distributions for
  any connectivity weight $\yxx$ conditional on the three generalized
  normal models defined in the text}
\label{fig:logit_function}
\end{figure}

The first two transformations, plotted in the upper panel of
\fig~\ref{fig:logit_function}, map the bounded domain $\opop{-1,1}$ of the
connectivity weights into the reals, and thus restrict the
generalized-normal likelihood to meaningful values of the connectivity
weights. The last model instead allows for connectivity weights outside
their meaningful bounds. It can be conceived as the model of a person who
has no precise knowledge of what the quantities $\yx$ are. Since the
clinician's final predictions concern health conditions given data $\yx$,
not the data $\yx$ themselves, this model can still be meaningfully used.
The probabilities for the connectivity weight $\yxx_i$ conditional on the
prior coefficients above are shown in the lower panel of
\fig~\ref{fig:logit_function}.

The prior coefficients are chosen by the following criteria: uniform
marginal distributions of the correlations between connectivity weights (leading
to $\ynuo=\yd+1$ as explained before); large uncertainty in the location
parameters ($\ykao = 1$); symmetry with respect to the origin ($\ymuo=0$);
a prior distribution for the connectivity as flat as possible (its second
derivative vanishes at the origin, leading to the values of $\yLao$ above).
In the case of the identity transformation we have chosen a $\yLao$ that
somewhat concentrates the prior around the true range of the connectivity
weights, $\clcl{-1,1}$.

The numerical values of the main quantities used throughout this study are
summarized in table~\ref{tab:numerical_values}.
\begin{table}[!ht]
  \centering
  \begin{tabular}{cl}
    $\ynh=55$& healthy patients\\
    $\yns=49$& schizophrenic patients\\
    $\yd=40$& graph parameters\\
    $\ykao, \ymuo, \ynuo, \yLao$& prior coefficients: see
                                  \eqns~\eqref{eq:logit_tr_consts_flat},
                                  \eqref{eq:tan_tr_consts_flat}, \eqref{eq:id_tr_consts_flat}
  \end{tabular}
  \caption{Numerical values in our study}
  \label{tab:numerical_values}
\end{table}

\subsection{Model comparison and selection}
\label{sec:model_comparison}

\subsubsection{Criteria for model comparison}
\label{sec:criteria}

Any two statistical models differ in two main characteristics: their
predictive power and their learning speed. Predictive power is a model's
capacity to give high probability to propositions that turn out to be true,
during and especially after its learning phase \citep[\cf][]{dawid1982b}.
Learning speed is how quickly a model reaches unchanging, stable predictive
probabilities as it gets updated with new data; note, however, that a model
may also never reach stable probabilities \citep[see \eg][]{bruno1964,berk1966};
\enquote{Alas, this seems like a model of the way things work in practical
  inference -- as more data comes in, one admits a richer and richer
  variety of explanatory hypothesis} \citep[\sect~3, p.~113]{diaconis1988}.
Thus, our choice of a model depends on the relative importance we give to
these two characteristics.

These two characteristics need not go hand in hand: a model can quickly
learn with very little data but settle on probabilities with poor
predictive power; conversely, it can reach great predictive power but only
after a long learning phase with a huge amount of data. A model can also
\enquote{unlearn}, \ie\ its predictive power initially increases and then
decreases before stabilizing. The latter phenomenon can happen because
every model initially makes its prediction through a mixture of likelihoods
-- the integral~\eqref{eq:goal_1st_exchangeability_parametric}, in our case
the t~distribution~\eqref{eq:predictive_distribution} -- but as the
training continues the mixture is replaced by a single likelihood, in our
case a generalized normal~\eqref{eq:generalized_normal}, as explained in
\sect~\ref{sec:generalized_normals}. It may happen that the mixture of
likelihoods has higher predictive power than a single likelihood, and in
this case a decline in predictive power will be observed. A model with such
behaviour is obviously unfit for the clinician's goal; this also means that
the sufficient statistics on which it is based capture very poorly the
differences in connectivity weights between health conditions.

When only a small amount of training data is available, it can be difficult
to assess which of two models has or will have the greater predictive
power. The first model can initially reach a greater predictive power than
the second, but the second model may eventually reach greater predictive
power than the first, with further training data. Models having the same
likelihood, however, have the same final predictive power; their learning
speeds depend on their parameter priors.

In a diagnostic problem like that faced by our clinician, the choice of a
model is ultimately dictated by the predictive power of the post-test
probabilities given by the model; but if we have little training data, the
learning speed of the model is also of some importance. Several
quantitative criteria can be conceived to assess these two characteristics:
\begin{enumerate}
\item\label{item:prob_all}the post-test probabilities the model gives to
  the correct health conditions for all training data, \ie\
  $\p(\yh_1,\yh_2,\dotsc)$;
\item\label{item:prob_fin}the post-test probabilities the model gives to
  the correct health conditions in the final phase of the training only,
  \ie\ $\p(\yh_{\text{last}} \| \yh_1,\yh_2,\dotsc)$;
\item\label{item:util_all}the expected utility the model yields for all
  training data;
\item\label{item:util_fin}the expected utility the model yields in the
  final phase of the training only.
\end{enumerate}
Post-test probabilities (criteria \ref{item:prob_all}, \ref{item:prob_fin})
are important for obvious reasons, and utilities (criteria \ref{item:util_all},
\ref{item:util_fin}) are important because the clinician's overall problem
is one of decision, as emphasized in the Introduction.
Consideration of all data (criteria \ref{item:prob_all},
\ref{item:util_all}) is important if we are interested in the performance
of the model for the whole set of patients; but consideration of the final
data only, conditional on the previous ones (criteria \ref{item:prob_fin},
\ref{item:util_fin}), tells us how much the model has learned
\citep[compare with a similar remark in model comparison using Bayes factors
by][]{bergeretal1996}.

We must keep in mind that these criteria assess a statistical model not by
itself but in combination with other factors, since they also depend on
pre-test probabilities (which can be influenced by other diagnostic tests)
or utilities.

Applied to our models, each of these four criteria gives a very similar
picture. We shall calculate the results for criteria~\ref{item:prob_fin}
and~\ref{item:util_fin}, the latter with two different utility tables.
This calculation can be explained in very intuitive terms:

Imagine that the $\ynh=55$ healthy and $\yns=49$ schizophrenic patients
visit the clinician in turn, in an unknown order. For each patient, let us 
further assume that the
clinician has pre-test probabilities $(\ypph,\ypps)=(0.5, 0.5)$, \ie\ she
is completely uncertain about the patient's health condition. The incidence
in the population is much lower than 50\%, of course, but the patients
presenting themselves for diagnosis are not representative of the full
population.

As stated in \sect~\ref{sec:exchangeability}, pre-test probabilities
represent the clinician's uncertainty before the fMRI test is made; they
can be based for example on a first diagnosis considering symptoms and
medical history of the patient and of the patient's family, on
psychological evaluations, and on other diagnostic tests. Here we assume
complete uncertainty for demonstration purposes.

The clinician acquires the fMRI result $\yx$ for that patient, and uses the
statistical model, trained with the data from all the patients that
previously visited her, to update to a post-test probability for
schizophrenia $\ypts$, given by \eqn~\eqref{eq:post-test_final}; obviously
$\ypth = 1-\ypts$. Now the clinician must make a decision -- say, treat or
dismiss -- based on the expected utilities of the decisions available. Each
decision has a different utility depending on the patient's true health
condition, as summarized by a table. We consider two tables: a symmetric
one
\begin{equation}
  \mbox{\setlength{\tabcolsep}{1ex}
  \begin{tabular}{lcc}
  (symmetric)  & healthy & schizophrenic\\\hline
    dismiss & $1$ & $0$ \\[-2\jot]
    treat & $0$ & $1$
  \end{tabular}}
  \label{eq:utility_table_symmetric}
\end{equation}
and an asymmetric one
\begin{equation}
  \mbox{\setlength{\tabcolsep}{1ex}
  \begin{tabular}{lcc}
  (asymmetric)  & healthy & schizophrenic\\\hline
    dismiss & $1$ & $-2$ \\[-2\jot]
    treat & $-1$ & $2$
  \end{tabular}}
  \label{eq:utility_table_asymmetric}
\end{equation}

In order to maximize the expected utility, as explained in
\sect~\ref{sec:decision_theory}, the clinician dismisses the patient if
$\ypts < 1/2$ in the case of the symmetric utility table, and if
$\ypts < 1/3$ in the case of the asymmetric
one; 
and treats the patient otherwise. After the clinician's decision is made,
the patient's true health condition is revealed and we record the actual
utility gained for each table, and the post-test probability the clinician
assigned to the true health condition, or its logarithm, usually called
\enquote{negative surprise} \citep{bartlett1952,good1956,good1957}. The
health condition and fMRI data of this patient are used to update the
model. The next patient is received, and so on, until all patients have
been examined.

The particular sequence of utilities and log-probabilities recorded in the
manner just described depends on the exact order by which the 104 patients
visit the clinician. We take an approximate average over all possible
$104! \approx 10^{166}$ orders by randomly sampling $520\,000$ of them. The
values of these averages for the final patient constitute the quantitative
criteria~\ref{item:prob_fin} and~\ref{item:util_fin}.

With the symmetric utility table~\eqref{eq:utility_table_symmetric}, the
average utility is also the average number of schizophrenic patients for
which the model yields $\ypts > 1/2$. The average utility for the last
patient, when the model has been trained with the rest of the patients, is
therefore a form of leave-one-out cross-validation \citep[vol.~2,
pp.~1454--1458]{allen1974,stone1974,kotzetal1982_r2006}. The asymmetric
table~\eqref{eq:utility_table_asymmetric}, slightly more realistic, tells
us that dismissing a schizophrenic patient has worse consequences than
treating a healthy one, and treating a schizophrenic patient has better
consequences than dismissing a healthy one \citep{mckenzie2014,hoetal2000}.
For this reason the patient is dismissed, more conservatively, only if
$\ypts<1/3$. Note that scaling a utility table by a positive factor or
shifting its values by a constant represent changes in the unit of measure
and in the zero of utilities, and therefore do not affect our relative
comparison of the statistical models.

\subsubsection{Results for our three models}
\label{sec:results_three_models}
\begin{figure}[!h]
  \centering
\includegraphics[width=0.75\linewidth]{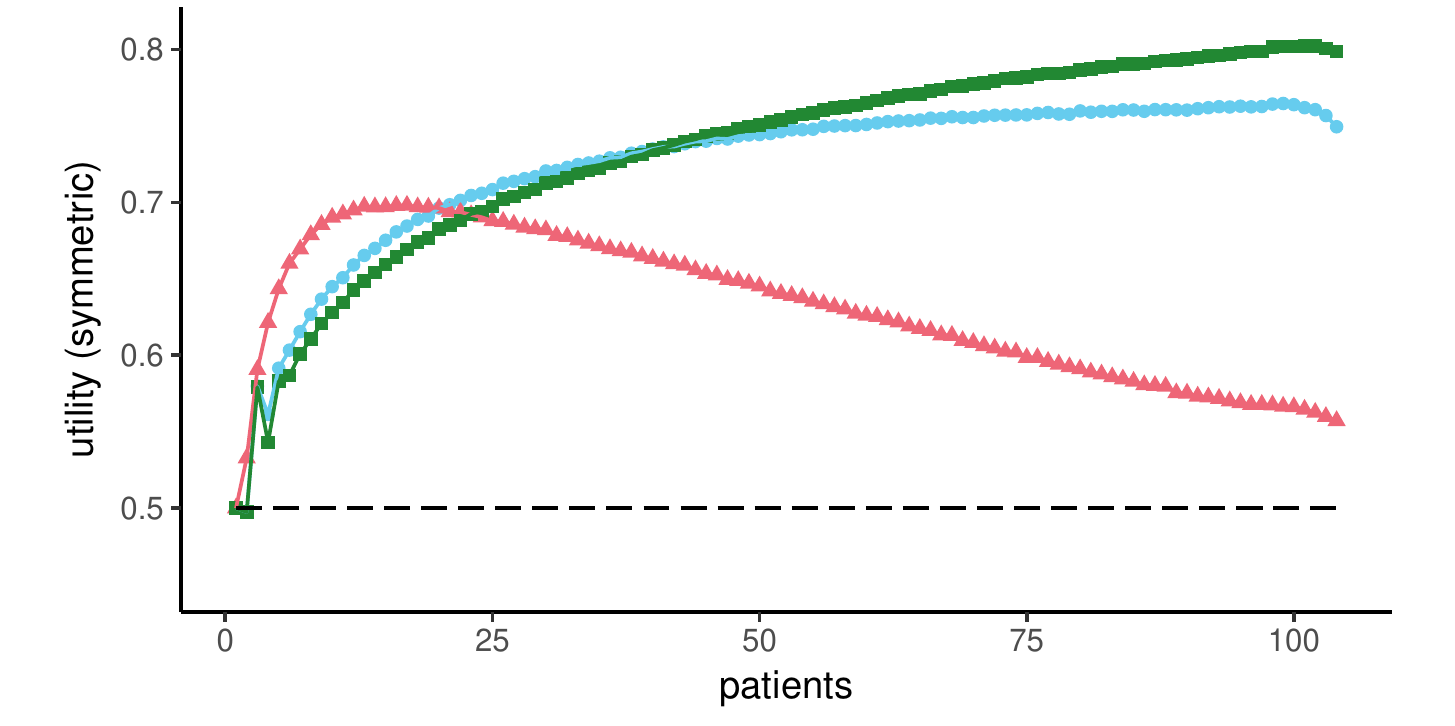}
\includegraphics[width=0.75\linewidth]{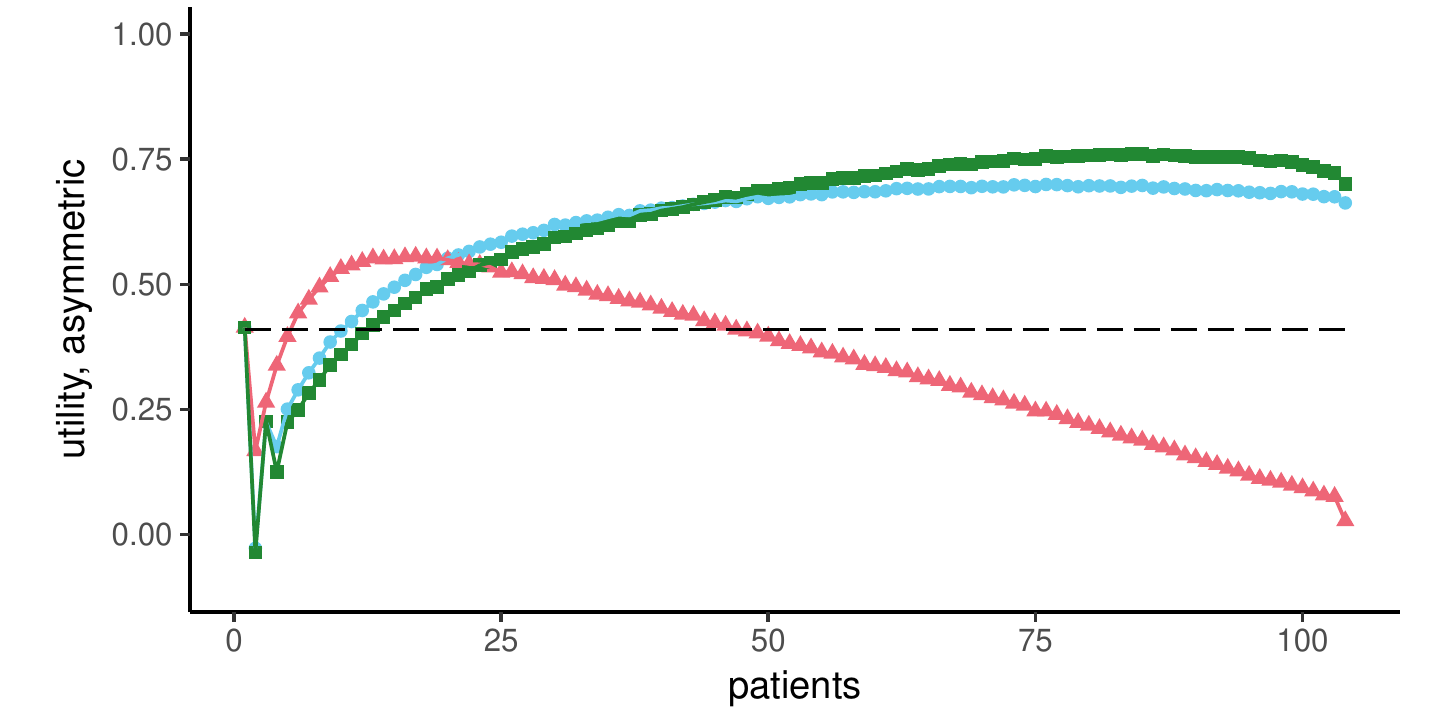}
\includegraphics[width=0.75\linewidth]{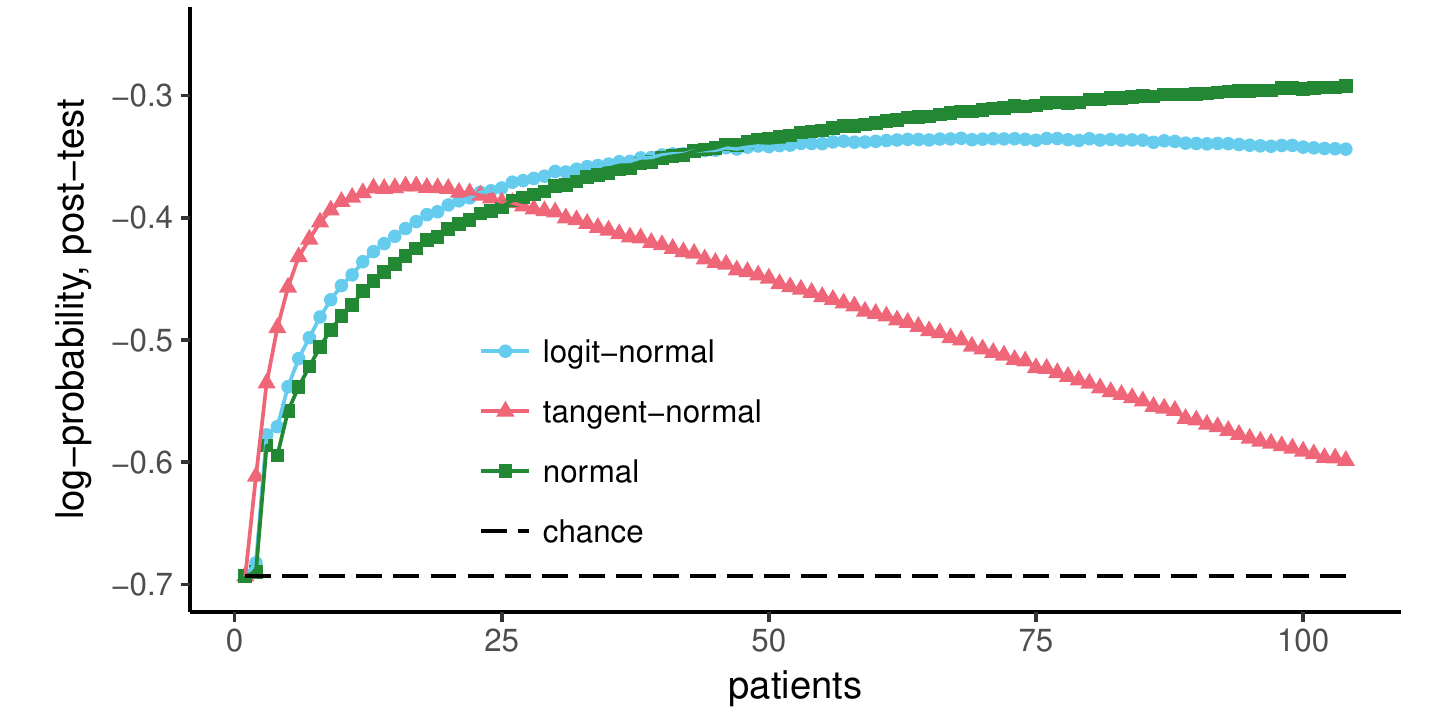}
\caption{Averaged sequence of utilities, with utility
  tables~\eqref{eq:utility_table_symmetric}
  and~\eqref{eq:utility_table_asymmetric}, and of log-probabilities for our
  three models. The standard deviations of the averages are smaller than
  the markers' size. The average values for the first and last patients are
  exact.}
\label{fig:utilities}
\end{figure}
The averaged sequences of utilities and log-probabilities calculated as in
\sect~\ref{sec:criteria} are shown in \fig~\ref{fig:utilities}. The R code
for the calculation is publicly available \citep{portamanaetal2017b}.
The results, summarized in table~\ref{table:results_models}, are
qualitatively identical by the two criteria and two utility tables we
chose: the normal model gives the best values for the final patient,
followed by the logit-normal model; the tangent-normal model has the worst
final predictive power, at or below chance level.
\begin{table}[!hb]
  \centering
  \mbox{\setlength{\tabcolsep}{1ex}
  \begin{tabular}{lccc}
    & symmetric utility~\eqref{eq:utility_table_symmetric}
    & asymmetric utility~\eqref{eq:utility_table_asymmetric}
    & log-probability \\\hline
    normal model  & $0.80$ & $0.70$ & $-0.29$\\
    logit-normal model  & $0.75$ & $0.66$ & $-0.34$\\
    tangent-normal model  & $0.56$ & $0.03$ & $-0.60$\\[0.5\jot]
    chance & $0.50$ & $0.41$ & $-0.69$ 
  \end{tabular}}
\caption{Final results for the three models}
  \label{table:results_models}
\end{table}

The plots of \fig~\ref{fig:utilities} illustrate the points made at the
beginning of \sect~\ref{sec:criteria}: one model can initially learn faster
than another and yet be overtaken in the later stages of learning; this is
the case for the logit-normal and normal models. The tangent-normal model
shows strong unlearning; this means that a tangent-normal likelihood and
its sufficient statistics do not distinguish well between healthy and
schizophrenic conditions. The slight downward bends at the final stages of
the logit-normal and normal models raise the suspicion that they might also
show some unlearning if further training data were supplied.

\begin{figure}[!t]
  \centering
\includegraphics[width=0.499\linewidth]{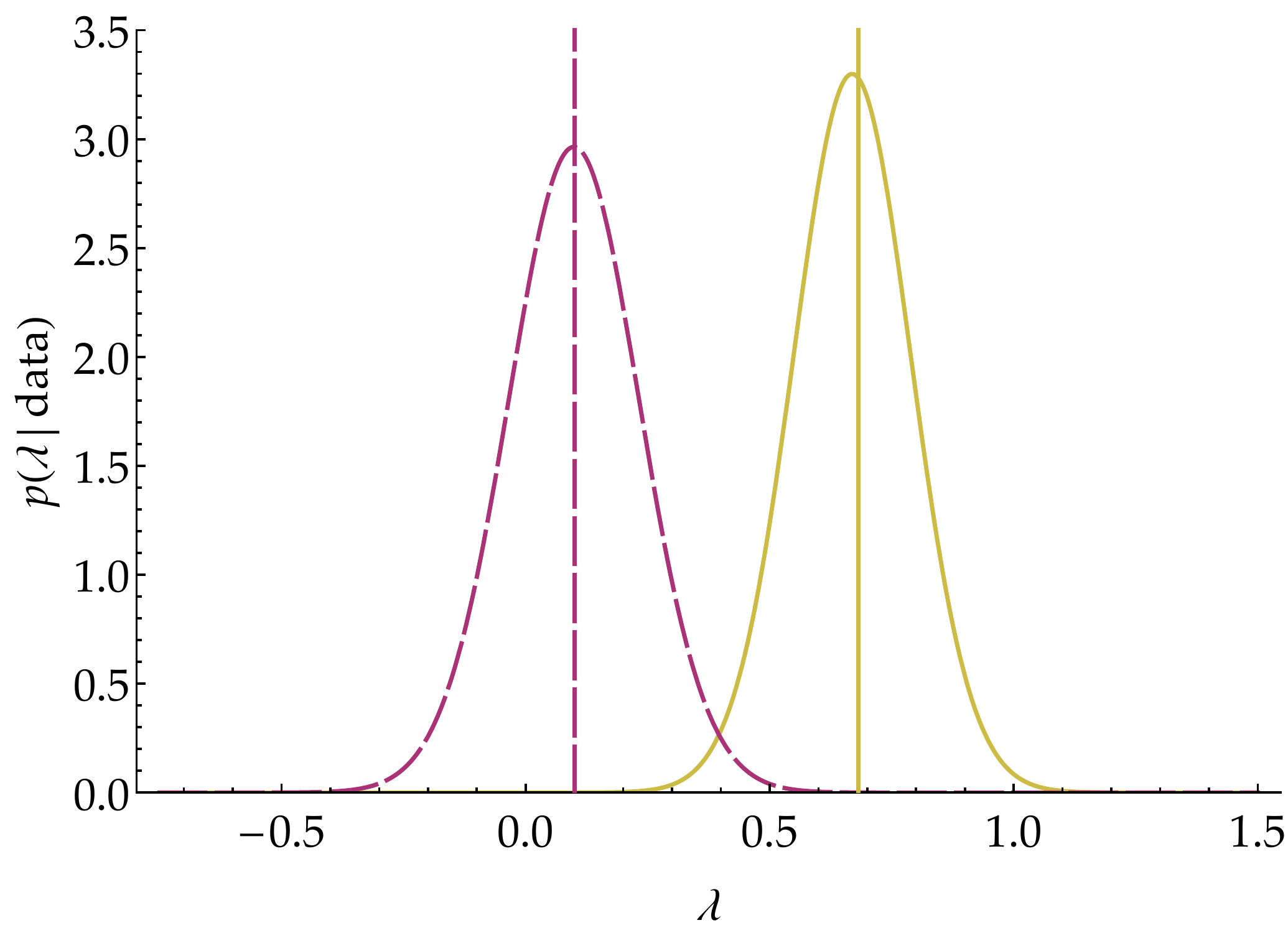}\hfill
\includegraphics[width=0.499\linewidth]{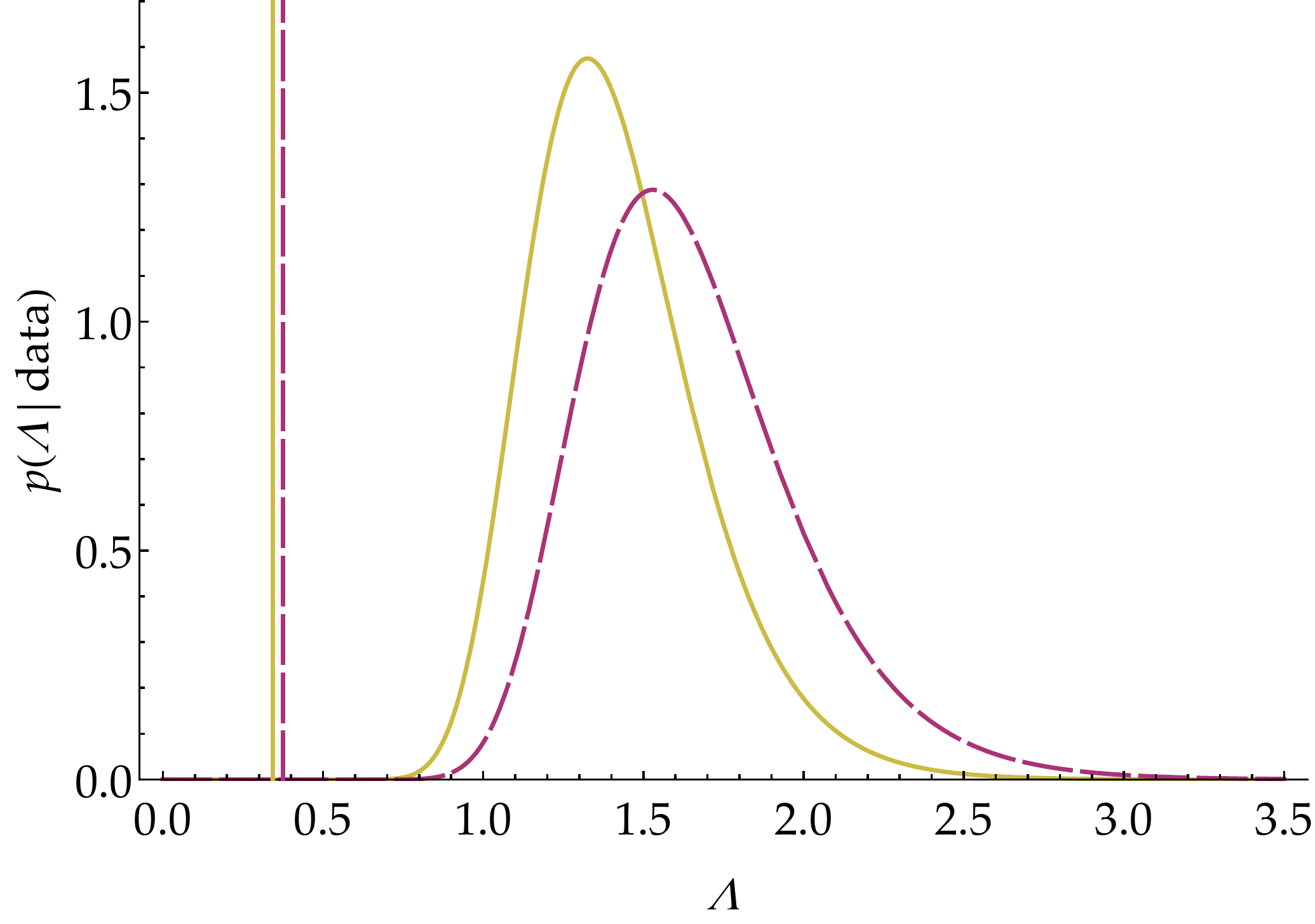}\\[\jot]
\includegraphics[width=0.499\linewidth]{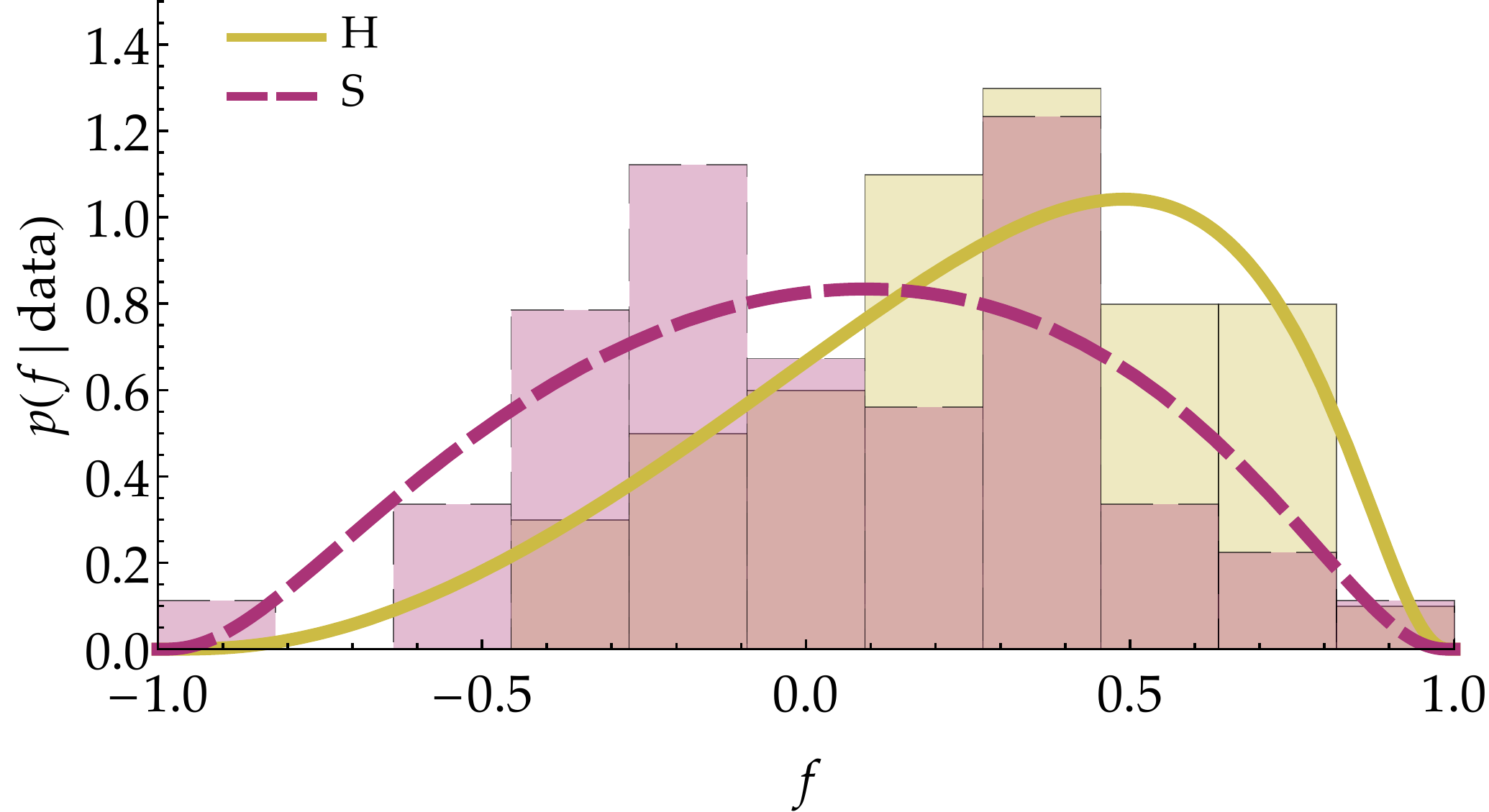}
\caption{Updated distributions of the logit-normal model for the location
  parameters $(\ylmm_{\yhu},\ylmm_{\yhd})$ (left), scale parameters
  $(\ylss_{\yhu}, \ylss_{\yhd})$ (right), and connectivity weights
  $\yxx_{\yhu}$, $\yxx_{\yhd}$ (bottom, superposed on the empirical
  distributions) for the connectivity between left superior parietal lobule
  and left lingual gyrus, corresponding to the bottom left
  panel of \fig~\ref{fig:overlapping_distribution}. 
  The vertical lines in the first two plots indicate the corresponding
  empirical statistics from the data.}
\label{fig:update_pars}
\end{figure}

The trends of the logit-normal and normal models suggest that the learning
phase is not finished: more patients are needed before their predictive
probabilities become stable. This is also evident from the updated marginal
distributions of their parameters
$(\ylmm_{\yhu},\ylss_{\yhu} ; \ylmm_{\yhd},\ylss_{\yhd})$, for example
those for the connectivity weight $\yxx$ between the left superior parietal
lobule and left lingual gyrus, shown in \fig~\ref{fig:update_pars} for the
logit-normal model. The distributions of the location parameters
$(\ylmm_{\yhu},\ylmm_{\yhd})$ have reached the empirical means of the data,
but those of the scale parameters $(\ylss_{\yhu}, \ylss_{\yhd})$ are still
very far away from the empirical variances. The reason is that the prior
for the scale parameters had a peak at a very large value of
$\yls\approx 20$. The $55$ data for healthy patients and $49$ for
schizophrenic ones have shifted this peak to $\yls_{\yhu} = 1.3$ and
$\yls_{\yhd} = 1.5$, but more data are needed to shift these peaks to even
smaller values -- provided that in the meantime the empirical values do not
change too much as new data are gathered.

The peak at high values of $\ylss$ is a known inconvenient feature of the
normal-inverse-Wishart conjugate prior, related to the correlation between
correlation and variance components of $\ylss$ characteristic of this prior
\citep[\eg][]{barnardetal2000}.

\subsubsection{Contrast with other model-comparison criteria}
\label{sec:variance_bayes_factors}

Common Bayesian model-comparison criteria are based on the joint
probability that the model gives to training data; especially its
logarithm, called \enquote{weight of evidence} or \enquote{marginal
  log-likelihood}, or the ratio of such logarithms, called Bayes factors
\citetext{\citealp[\chaps~V, VI, A]{jeffreys1939_r2003};
  \citealp{good1950,mackay1992,kassetal1995}}. The simple reason is Bayes' theorem:
\begin{equation}
  \label{eq:explain_model_probability}
  \p(\text{model} \Cond \text{data} \Land \text{prior info} ) \propto
  \p(\text{data} \Cond \text{model} \Land \text{prior info}) \times
  \p(\text{model} \Cond \text{prior info} ),
\end{equation}
the latter probability usually assumed the same for all models \citep[but
see][]{portamana2017c}. A higher weight of evidence means that the model is
more probable.

In our study, however, we have two kinds of data: health conditions and
fMRI results. Since the likelihood for the health condition, used by the
clinician to arrive at a post-test probability, gives the probability for
the fMRI results $\set{\yx_i}$, it seems intuitive to calculate the weights
of evidence of our models based on these data. The result is the opposite
of what we obtain with the averaged-utility criterion or any of other three
mentioned above. We obtain:
\begin{equation}
  \label{eq:weights_evidence}
  \begin{aligned}
    &\ln\pf(\text{fMRI results} \Cond \text{logit-normal model} \Land \text{health conditions}) = -1913,\\
    &\ln\pf(\text{fMRI results} \Cond \text{tangent-normal model} \Land \text{health conditions}) = -1858,\\
    &\ln\pf(\text{fMRI results} \Cond \text{normal model} \Land \text{health conditions}) = -2488,
  \end{aligned}
\end{equation}
which gives the normal model a much smaller probability than the other two,
and the tangent-normal model the highest.

This discrepancy with the averaged-utility criterion is not completely
surprising, though. Imagine a disease that leads to no differences at all
between the fMRI results of patients with the disease and those of healthy
controls. If we found a statistical model that predicted the fMRI results with
certainty, this model would thus have a the highest weight of evidence
(zero), and yet its final average utility would be at chance level, since
it could not help us at all in telling healthy from diseased patients. For
our problem the right comparison and selection criterion is the utility or
one of the other three criteria previously listed.

\subsubsection{Final assessment of models}
\label{sec:assessment}

\begin{figure}[!h]
  \centering
\includegraphics[width=0.75\linewidth]{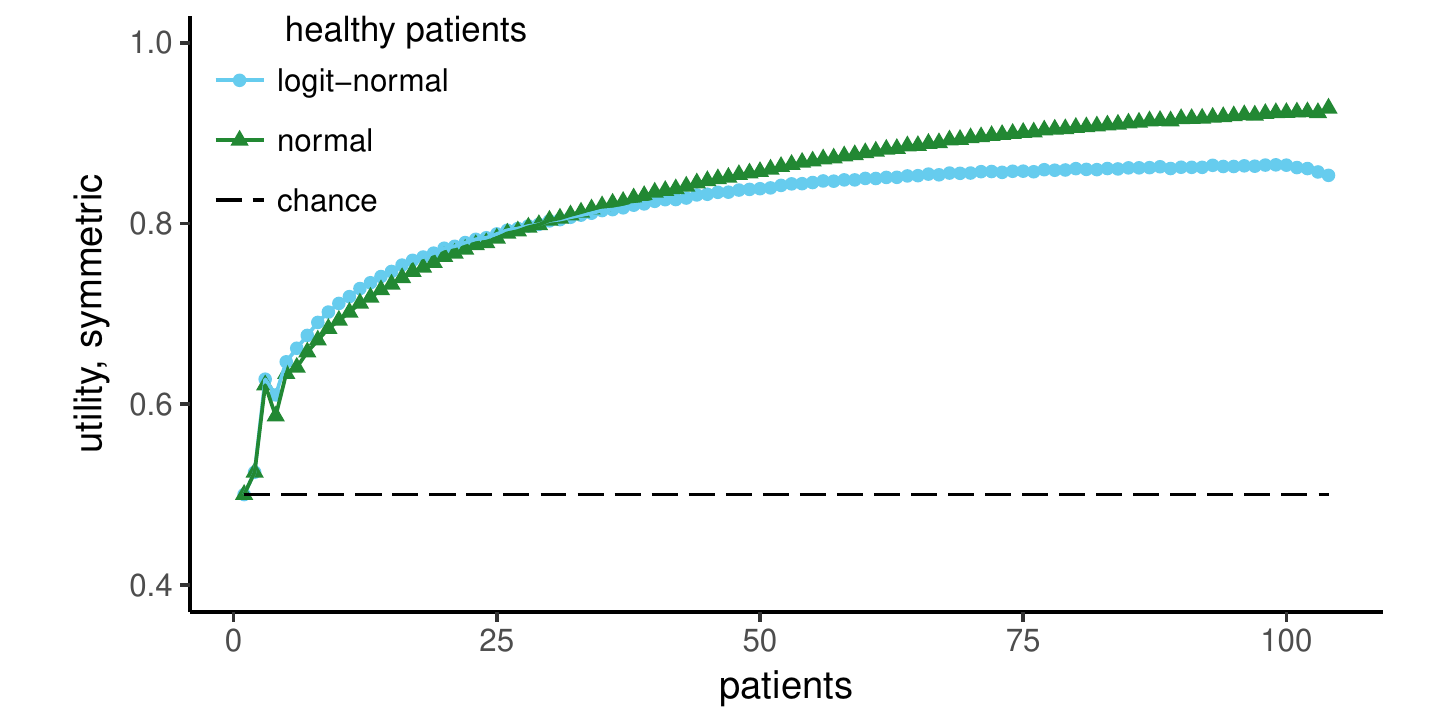}\\
\includegraphics[width=0.75\linewidth]{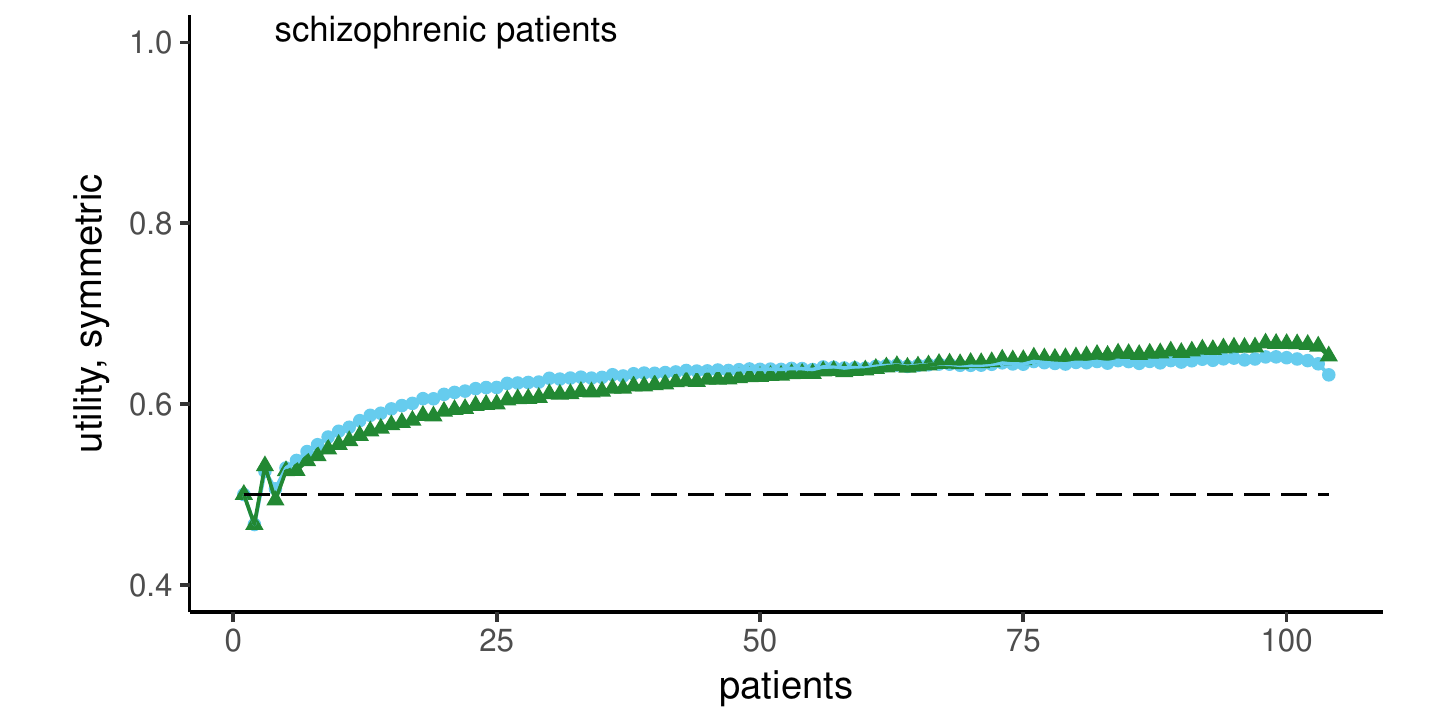}
\caption{Sequence utilities averaged over healthy (left) and schizophrenic
  (right) patients separately, for the logit-normal and normal models}
\label{fig:utilities_separate}
\end{figure}

The average-utility criterion clearly excludes the tangent-normal model,
which even shows a rapid unlearning. The logit-normal and normal models
have almost similar performances, at around $75$--$80\%$ of final patients
correctly treated. We can also plot the sequence of utilities averaged over
healthy and schizophrenic patients separately, as in
\fig~\ref{fig:utilities_separate}, which gives us an idea of the ratio
between true and false negatives, and true and false positives. Both models
show around $35\%$ false positives (dismissed schizophrenic patients), with
the normal model giving slightly higher final rates of true negatives,
$93\%$, and true positives, $65\%$, than the normal, $85\%$ and $63\%$.

However, we emphasize that the features used as inputs to the model were selected using a very simple heuristic (maximum difference in means between the two groups, see \sect~\ref{sec:data_reduction}), and the statistical model was selected for its computational properties rather than its fit to the distributions of connection weights derived from fMRI data.
This notwithstanding, we believe that with further training, the logit-normal model could
reach a higher predictive power. The reason is that some empirical
distributions of connectivity weights, like the one for schizophrenic
patients shown in red in \fig~\ref{fig:update_pars}, seem to be bimodal;
and the logit-normal likelihood, unlike the normal one, is capable of
bimodality, thus fitting these distributions better.

Our assessment, however, is just an illustrative example for the general
method discussed in this paper, and we are not earnestly proposing the
logit-normal model (nor any other specific model) as the optimal one to use in the
problem of diagnosing schizophrenia. We note that any model assessment and selection using the average-utility metric depends on several important quantities and
assumptions:
\begin{enumerate}[label=\Alph*.]
\item the pre-test probabilities given by the clinician; we assumed these
  to be $(0.5, 0.5)$;
\item the clinician's range of decisions; we assumed it to be simply
  \enquote{treat or dismiss}, but it could comprise several different
  kinds of treatments;
\item the utilities of the  clinician's decisions; we assumed these to be as
  in formula~\eqref{eq:utility_table};
\item the ratio between the numbers of healthy and schizophrenic training
  data; $55/49=1.12$ in our case;
\item\label{item:order_important}the clinician diagnoses one patient at a time.
\end{enumerate}

For a proper model assessment we should therefore investigate and consider
more realistic rates of healthy vs schizophrenic cases that visit a
particular clinician, in order to have better-informed pre-test
probabilities; and we should consider more realistic decisions available to
the clinician, together with realistically quantified utilities.

\medskip

Assumption \ref{item:order_important} deserves some explanation as it may
mistakenly appear that it doesn't matter whether patients visit the
clinician simultaneously or one at a time. Suppose two patients, Tom and
Joe, visit the clinician together, and the clinician obtains fMRI data for
both, $\yx_{\tom}$ and $\yx_{\joe}$. The joint post-test probability for
their health conditions $\yh_{\tom}$ and $\yh_{\joe}$ is different from the
one obtained first calculating Joe's one, say, and then Tom's using Joe's results:
\begin{equation}
  \label{eq:tom_joe_relevance}
  \pf(\yh_{\tom},\yh_{\joe} \| \yx_{\tom}, \yx_{\joe}, \yD, \yMl) \ne
  \pf(\yh_{\tom} \| \yx_{\tom}, \yh_{\joe},\yx_{\joe}, \yD, \yMl) \times
  \pf(\yh_{\joe} \|\yx_{\joe}, \yD, \yMl).
\end{equation}
This inequality can be easily verified by applying the probability product
rule to the left side, and can be understood as follows. Suppose the
clinician first wants to calculate the likelihood for Joe's being healthy.
If Tom is schizophrenic then his fMRI result is unimportant for Joe's
likelihood, owing to our assumption of independent
priors~\eqref{eq:logit-normal_parameters_factor_condition}. But if Tom is
healthy, then his fMRI result, \emph{which is known to the clinician},
should lead to an updated model for Joe's likelihood. The likelihood for
Joe's being healthy is therefore a mixture of these two possible
likelihoods, with weights proportional to the post-test probability for
Tom's health condition. Thus, Joe's likelihood is affected by Tom's fMRI
result even if Joe's is calculated first and Tom's health condition is not
yet known. More generally, if several patients visit the clinician
simultaneously, she should order diagnostic fMRI tests for all of them at
once and calculate a joint post-test probability for them, in order to make
the best-informed prediction for each.

\section{Discussion}
\label{sec:discussion}

\subsection{Summary}
\label{sec:summary}

The diagnosis of a medical condition is a complex process that takes in a
variety of judgements and evidence from the clinician and from any
diagnostic tests available to her. Bayesian probability theory has found
wider acceptance in medicine because it can consistently combine and frame
such judgements and evidence
\citep{goodman1999,davidoff1999,greenland1998}.
Formulae~\eqref{eq:pre-test_prob_intro}--\eqref{eq:test_prob_intro} show
the basic scheme of how the clinicians' judgements and the results of
diagnostic tests are combined \citep{soxetal1988_r2013}. The role of a
diagnostic test is not simply to give a dichotomic answer, \eg\
healthy/ill, but a \emph{likelihood} for each health condition, to be
combined into this scheme together with the likelihoods from other tests.
The final probability obtained from these likelihoods is finally used by
the clinician to decide upon a course of action, \eg\ dismiss/treat
(\sect~\ref{sec:decision_theory}).

The values of likelihoods from such tests need to be determined from a set
of training set of data for each health condition. In this work we have
discussed how to determine the likelihoods when the diagnostic test and
training set consist of fMRI data, considering for concreteness the case of
schizophrenia as the disease in question, and by using real fMRI data from
healthy and schizophrenic patients (\sect~\ref{sec:data_acquisition}). We
derived them step by step from first principles through a sequence of
assumptions:
\begin{enumerate}[(a)]
\item\label{item:as_exchang}\textbf{Partial exchangeability} with respect
  to the health conditions, explained in \sect~\ref{sec:exchangeability}.
  We believe this assumption to be very natural in medical diagnostics. By
  itself it already leads to a specific expression for the likelihood,
  \eqn~\eqref{eq:goal_1st_exchangeability}, although this expression is
  very difficult to compute.
\item\label{item:as_suff}\textbf{Sufficiency} of an empirical statistics of
  a reduced set of fMRI data
  (\sects~\ref{sec:data_reduction}--\ref{sec:sufficiency}). We believe such
  kind of assumption to be sound and at least approximately true when
  neurologically motivated, and moreover it provides a bridge between
  biophysical considerations and the specification of probabilities. It
  leads to a likelihood,
  \eqn~\eqref{eq:goal_1st_exchangeability_parametric}, amenable to
  numerical or analytic computation.
\item\label{item:as_prior}\textbf{Prior knowledge} of the empirical
  statistics for the different health conditions
  (\sect~\ref{sec:our_models}). An assumption of this kind is always
  necessary, especially with small training data sets; but it affects our
  inferences less and less as our training data accumulate.
\end{enumerate}
In our study, for~\ref{item:as_suff} we specifically assumed the
sufficiency of the first and second moments of some functions
(\sect~\ref{sec:sufficiency}) of the functional-connectivity weights
obtained from the fMRI data (\sect~\ref{sec:data_reduction}). Our focus on
connectivities was neurologically motivated, but our choice of first and
second moments of particular functions was made for mathematical
simplicity, in order to illustrate the method. Our choice of roughly flat,
conjugate priors for assumption~\ref{item:as_prior} was also made for the sake of
mathematical simplicity and illustration (\sect~\ref{sec:our_models}).

Notwithstanding our simple, mainly illustrative choices, we obtained good
diagnostic performances, briefly discussed in the next section. To assess
these performances and the relative predictive power of different choices
in assumption~\ref{item:as_suff} above, we presented several criteria,
based on decision theory (\sect~\ref{sec:model_comparison}). These criteria
also help in understanding whether the training of the likelihoods has
stabilized (\sect~\ref{sec:model_comparison}). We observed that in this
kind of study decision-theoretical criteria can yield results in seeming
contrast with Bayesian model-comparison criteria, like weights of evidence
and Bayes factors (\sect~\ref{sec:variance_bayes_factors}); but this
contrast is understandable and is not a sign of inconsistency.

We emphasize that the analysis and conclusions presented here are just
illustrations of the general method, and are not meant to be used for real
diagnoses: as explained in \sects~\ref{sec:suff_stat} and
\ref{sec:assessment}, for a real application we would first need to investigate
better-informed sufficient statistics, realistic decisions available to the
clinician, the latter's utilities, and realistic pre-test probabilities. In
this study we used simple default values for illustrative purposes only.

\subsection{Comparison with other studies and methods}
\label{sec:comparison_other_studies}

In this study we used a rather naive approach, explained in
\sect~\ref{sec:data_reduction}, to select a subset of brain connections for
our analysis. That approach can be criticized in two different ways. First,
it ignores the fact that if distributions are narrow, the means can be
close together without much overlap, and that such distributions are likely
to be more informative for the purposes of discrimination. This could be
improved, for example, by taking the minimal area of the distributions'
overlap (dark red area in \fig~\ref{fig:overlapping_distribution}) instead.
Second, we did not restrict our search for suitable connections and
associated brain areas to the areas that are known to be part of
resting-state networks identified in previous studies, like the default
mode network. Our lack of specificity, though, was motivated by previous
studies which have demonstrated that functional connectivity can be also
measured in other task-related networks, induced by spontaneous activity
\citep{He2009}, and that in resting state different activity patterns can
appear in schizophrenic patients, owing \eg\ to hallucinations during the
scan \citep[\eg][]{Shergill2000}.

Despite our simplified and possibly unrealistic choices of connectivity
weights, sufficient statistics, parameter priors, pre-test probabilities,
and utility functions, we obtained a diagnostic performance of $80\%$,
measured by leave-one-out cross-validation
(\sect~\ref{sec:model_comparison}), comparable to classification results
based on fMRI data reported in previous studies
\citetext{\eg, \citealp[18~healthy + 18~schizophrenic patients]{venkataramanetal2012}; 
  \citealp[45~+ 46 patients]{cetinetal2016}; 
  \citealp[36~+ 34 patients]{demircietal2008}
}.

\medskip

Our results also compare with classification results that we achieved using
machine-learning methods. Cross-validation tests using support-vector
machines (using $80\%$ of the data for training and $20\%$ for testing in a
randomized iterative way) also yielded around $80\%$ of correct diagnoses (data not shown).

But, as explained in the Introduction (points~\ref{item:test_combination},
\ref{item:no_classif}), methods like these, which simply classify or are
deterministic, do not fit the clinician's decision-theoretical problem:
they cannot be combined with other diagnostic tests and do not fit a
general decision-theoretic approach -- \cf\ \sects~\ref{sec:criteria},
\ref{sec:decision_theory}. Most machine-learning methods
\citep{bishop2006,murphy2012} are thus ruled out.

There is no real contrast, however, between machine-learning methods and
the method presented here: machine-learning algorithms can be interpreted
as convenient, fast approximations  of
Bayesian methods \citep[see \eg\ the explicative image in][]{huszar2017},
often combined with default utility functions and decision rules
\citep{murphy2012,mackay1995_r2003,mackay1992,mackay1992b,mackay1992c,mackay1992d,mackay1992e}.
A machine-learning classification algorithm that gives a good performance
can suggest good likelihoods or parameter priors to be used in a
statistical model. For example, the simplest version of a support-vector
machine \citetext{\citealp[\chap~7]{bishop2006};
  \citealp[\sect~14.5]{murphy2012}} can be interpreted as a model where the
likelihood $\yL_{\yhu}(\yx\|\ythh)$ for one health condition is very large
in a half of the dataspace $\yx\in\clcl{-1,1}^{\yd}$ and zero in the rest,
the hyperplane separating these half-spaces being determined by the
parameter $\ythh$. The likelihood $\yL_{\yhd}(\yx\|\yths)$ for the other
health condition is likewise large or zero in two half-spaces separated by
a hyperplane determined by $\yths$; see
\eqns~\eqref{eq:goal_1st_exchangeability_parametric} in
\sect~\ref{sec:parametric_models}. In this model the parameter prior
$\pf(\ythh, \yths \|\yM ,\yI)$ correlates the two parameters
$(\ythh, \yths)$ in such a way that the two hyperplanes coincide and the
two likelihoods have support on opposite sides. As the model is trained and
the parameter prior is updated, \eqn~\eqref{eq:updated_hyperprior}, the
hyperplane moves in space in a way to maximize the product of the
likelihoods. This corresponds to the search of an optimal separation
hyperplane by the support-vector machine.

\subsection{Possible improvements}
\label{sec:improvements}

Besides using more realistic pre-test probabilities, and utility values,
the method presented here could be improved in several other respects,
especially with regard to assumptions~\ref{item:as_suff}
and~\ref{item:as_prior} summarized in \sect~\ref{sec:summary}.

In point~\ref{item:as_suff} we assumed that particular connectivity weights
are sufficient to distinguish among schizophrenic and healthy patients.
These connectivities were calculated as in \sects~\ref{sec:data_reduction}
and~\ref{sec:ROI_selelection_and_conn_measure}. More sophisticated choices
of Regions Of Interests and functional-connectivity measures
\citep{marrelecetal2011,Smith2011,Wang2014,gheiratmandetal2017,demircietal2008}
or even integration of functional and structural imaging
\citep{michaeletal2010} could obviously lead to an increased predictive
power. A different choice of sufficient statistics could also improve the
performance. With increased computational power it would even be possible
to use the full set of connectivities rather than sufficient statistics
thereof -- so-called \enquote{nonparametric} models
\citep[\cf][]{zhangetal2014b,zhangetal2016,nielsenetal2016,kooketal2017}.

Regarding assumption~\ref{item:as_prior}, in
\sect~\ref{sec:generalized_normals} we mentioned that parametric models may
lead to probabilities that lack a closed form and are analytically
intractable. The models we chose, with conjugate prior, have the advantage
of having closed-form formulae, but they also restrict our choice of
sufficient statistics and parameter priors. Higher predictive power could
be achieved by using other kinds of sufficient statistics, leading to
likelihoods that are not generalized normals, or by using non-conjugate
priors, \eg\ treating means, correlations, and variances independently, as
discussed by \cite{barnardetal2000}. In this case numerical methods are
needed, such as Markov-Chain Monte Carlo sampling and integration
\citetext{\citealp[\chap~IV]{mackay1995_r2003};
  \citealp[\chaps~23--24]{murphy2012}}. It must be kept in mind, though,
that numerical methods can be computationally vastly more expensive than
analytic ones. In preliminary studies that led to our present work we
considered a couple of statistical models that require Monte Carlo
sampling: a truncated normal and a product of beta distributions among
them. The calculation of the relevant integrals for these models has vastly
higher time costs than for the closed-form models presented here. For
example, calculation of a posterior parameter distribution for the
truncated-normal model required $17\:\mathrm{h}$ on a computer cluster;
whereas the corresponding calculation for the models in the present work
takes a fraction of a second on a laptop. To assess and select a model for
our clinician to use, such integrals need to be computed over and over, as
we explained in \sect~\ref{sec:criteria}. The assessment of statistical
models that require Monte Carlo methods can therefore lead to months of
computation. Further explorations are needed in this direction.

The comparison with support-vector machines sketched at the end of the
previous section shows one important assumption of our statistical models:
the independence of the prior parameter distributions for the two health
conditions:
$\pf(\ythh, \yths \|\yI) = \pf(\ythh \|\yI) \times \pf( \yths \|\yI)$, see
\eqn~\eqref{eq:logit-normal_parameters_factor_condition}. Statistical
models of which support-vector machines are approximations clearly cannot
make this assumption. The performance of the model might improve by using a
non-independent joint prior distribution, which allows training data for
one health condition to influence the parameter distribution for the other.
In the case of the generalized-normal models we examined, this can be
achieved -- whilst preserving their computational convenience -- by taking
a weighted average of several values of prior
coefficients~\eqref{eq:logit-normal_parameters}. This is known as a
hierarchical model \citep[\sect~4.6.5]{bernardoetal1994_r2000}.

The possibilities for improvement listed in these last paragraphs suggest that
the method we have presented here, hinging on first principles, has great
potential for applications and for development in different directions; moreover this is not to the exclusion of other methods, but assimilating
their principles and advantages.

\section{Methods}
 
\subsection{Data preprocessing}
\label{sec:ROI_selelection_and_conn_measure}

Preprocessing of the rfMRI images is carried out using the \textsc{fmrib}
Software Library tools \citep[\textsc{fsl},
v5.08:][]{Jenkinson2012,Smith2004} and consists of the following steps:
removal of the first ten image volumes, leaving the remaining 130 volumes
for further data processing; removing non-brain tissue
\citep[\textsc{Bet}:][]{smith2002c}; motion correction
\citep[\textsc{mcflirt}:][]{Jenkinson2002}; spatial smoothing with a
$6\;\mathrm{mm}$ full width at half maximum normal kernel; temporal
low-pass filtering with a cut-off frequency of $0.009\;\mathrm{Hz}$; white
matter and cerebrospinal fluid regression (\textsc{fsl} regfilt,
\textsc{melodic}).

\medskip

For each subject we first linearly register the rfMRI image first to the
structural, skull-removed image (image segmentation for skull removing with
\textsc{spm8}, Wellcome Department of Cognitive Neurology, London,
\textsc{ukfsl}; linear registration with \textsc{fsl}/\textsc{flirt}:
\citealt{Jenkinson2001,Jenkinson2002}) and then, through a non-linear
mapping, to the MNI standard brain (non-linear registration with Advanced
Normalization Tools \citep[\textsc{ANTs}:][]{Avants2011}; MNI 152 standard
brain, non-linear 6th generation \citep{Grabner2006}. Regions of interest
(ROIs) of the resulting functional image in standard space are extracted
such that they match the 94 regions identified by the Oxford lateral
cortical atlas with a probability above $50\%$ \citep{Desikan2006}. The
temporal mean signals across the voxels in each ROI are used to calculate
the functional connectivity measured based on the Pearson correlation
coefficient.

\subsection{The normal model with conjugate prior}
\label{sec:normal-inverse-wishart}

This statistical model, denoted in this section by $\yM$, is amply
discussed in the literature
\citetext{\citealp[\sect~3.6]{gelmanetal1995_r2014};
  \citealp{minka1998_r2001,murphy2007}}; here we only give a summary.

Its likelihood is a normal distribution
\begin{equation}
  \label{eq:likelihood_normal}
  \yL(\yx \| \ylmm, \ylss, \yM, \yI)
  =
  \dnormal[\ytr(\yx) \| \ylmm, \ylss]\, \ytr'(\yx)
\end{equation}
with mean $\ylmm$ and covariance matrix $\ylss$.

The prior distribution for the parameters $(\ylmm, \ylss)$ is a
normal-inverse-Wishart distribution, \ie\ the product of a normal
distribution for $\ylmm$ and an inverse-Wishart matrix distribution \citetext{\citealp[\sect~3.4]{guptaetal2000}; \citealp{tiaoetal1964};
  \citealp[\sect~3.2.5]{bernardoetal1994_r2000}} for $\ylss$:
\begin{gather}
  \label{eq:logit-normal_parameters_factor_individual}
  \pf(\ylmm, \ylss \|\ykao,\ymuo,\ynuo,\yLao,\yMl, \yI) =
  \pf( \ylmm\| \ylss, \ykao,\ymuo, \yMl, \yI)
  \times \pf( \ylss\| \ynuo,\yLao, \yMl, \yI),
  \\[\jot]
  \shortintertext{with}
  \begin{aligned}
    \label{eq:logit-normal_parameters}
    \pf( \ylmm\| \ylss, \ykao, \ymuo, \yMl, \yI)
    &= \dnormal(\ylmm \| \ymuo, \ylss/\ykao),
    \\
    \pf( \ylss\| \ynuo,\yLao, \yMl, \yI)
    &= \diwishart(\ylss \| \ynuo, \yLao)\propto
      \det(\ylss)^{-\frac{\ynuo+\yd+1}{2}}
      \,
      \exp\bigl(-\tfrac{1}{2} \tr\yLao\ylss^{-1}\bigr).
  \end{aligned}
\end{gather}
It should be noted how $\ylss$ appears as parameter in the distribution for
$\ylmm$, so their distributions are not independent. The composite
distribution depends on two scalar, one vector, and one matrix coefficients
$(\ykao,\ymuo,\ynuo,\yLao)$.

This prior parameter distribution retains the same form when it is
conditioned on the data $(\yx_i)$ of $\yn$ patients, becoming a posterior
parameter distribution with updated coefficients $(\yka,\ymu,\ynu,\yLa)$
depending on the prior ones and on the sufficient statistics:
\begin{equation}
  \label{eq:update_coeffs}
    \begin{aligned}
    \yka &= \ykao + \yn,\quad     &\ynu &= \ynuo + \yn,\\
    \ymu &= \frac{\ykao\,\ymuo+\yn\, \av{\yx}}{\ykao+\yn},\quad
    &\yLa &= \yLao + \yn\, \Cov(\yx)
           + \frac{\ykao\,\yn}{\ykao+\yn} \bigl(\av{\yx}-\ymuo\bigr) \bigl(\av{\yx}-\ymuo\bigr)\T.
  \end{aligned}
\end{equation}

The main features of the normal-inverse-Wishart distribution for
$(\ylmm, \ylss)$ are these:
\begin{equation}
  \label{eq:moments_modes_norm-iwishart}
  \begin{aligned}
    &\text{$\ylmm$:}\quad\;\hphantom{\Biggl\{}\text{mean \amp\ mode} = \ymu,
      \qquad\text{covariances} = \frac{1}{\yka\,(\ynu-\yd-1)}\yLa;\\
    &\text{$\ylss$:}\quad\left\{ 
      \!\begin{aligned}
          &\text{mean} = \frac{1}{\ynu-\yd-1}\yLa,
            \qquad\text{mode} = \frac{1}{\ynu+\yd+2}\yLa,\\
          &\text{diagonal variances} =
            \frac{2}{(\ynu-\yd-3)(\ynu-\yd-1)^2}{(\yLa)_{kk}}^2.
      \end{aligned}\right.
  \end{aligned}
\end{equation}
These formulae above say that the uncertainty in the location parameter
$\ylmm$ decreases as $\yka$ and $\ynu$ increase for fixed $\yLa$, and the
uncertainty in the matrix scale parameter $\ylss$ decreases with increasing
$\ynu$. When $\ynu=\yd+1$ the marginal distributions for the correlations
of $\ylmm$ are uniform \citetext{\citealp[\sect~3.6]{gelmanetal1995_r2014};
  \citealp[\sect~2.2]{barnardetal2000}}. Because of these properties, a
\enquote{vaguely informative} parameter distribution should have small
$\ykao$ and $\ynuo$ 
\citep{minka1998_r2001,murphy2007}.

When the likelihood~\eqref{eq:likelihood_normal} and the parameter
prior~\eqref{eq:logit-normal_parameters_factor_individual}, updated with
\eqref{eq:update_coeffs}, are multiplied and the parameters are integrated,
the resulting distribution for $\yx$ is a multivariate t~distribution
\citep{kotzetal2004,minka1998_r2001,murphy2007}
\begin{equation}
  \label{eq:predictive_distribution}
  \pf[\yx \| (\yx_i),\ykao,\ymuo,\ynuo,\yLao,\yMl, \yI] \equiv
  \pf(\yx \| \yka,\ymu,\ynu,\yLa,\yMl, \yI) =
  \dstudentt\Bigl[\ytr(\yx) \bigcond \ynu-\yd+1, \ymu,
  \tfrac{\yka+1}{\yka\,(\ynu-\yd+1)}\yLa \Bigr]
\end{equation}
with $\ynu-\yd+1$ degrees of freedom, mean $\ymu$, scale matrix
$\frac{\yka+1}{\yka\,(\ynu-\yd+1)}\yLa$, and covariance matrix
$\frac{\yka+1}{\yka\,(\ynu-\yd-1)}\yLa$.

\subsection{Decision theory  and utility}
\label{sec:decision_theory}

Once we have the post-test probabilities $(\ypth,\ypts)$ for the possible
health conditions of a patient given the fMRI data, there remains to decide
upon a course of action. This is the domain of decision theory
\citetext{\citealp{raiffaetal1961_r2000}; \citealp[\chaps~13,
  14]{jaynes1994_r2003}; \citealp[\chaps~6, 7]{soxetal1988_r2013}}.

Suppose we have only two courses of action: treat $\yT$ or dismiss $\ynT$.
A decision-theoretical analysis needs, besides the probabilities for the
health conditions, also the utilities (or costs) of choosing an action
given the patient's true health condition. For example, treatment of a
healthy patient could harm the latter, or it could be innocuous. With two
courses of action and two health conditions we have four utilities
$\yu_{\text{decision}\|\text{condition}}$:
\begin{equation}
  \mbox{\setlength{\tabcolsep}{1ex}
  \begin{tabular}{lcc}
    & healthy & schizophrenic\\\hline
    dismiss & $\yudh$ & $\yuds$ \\[-2\jot]
    treat & $\yuth$ & $\yuts$
  \end{tabular}}
\qquad
\begin{aligned}[t]
\yudh&>\yuth,\\ \yuts&>\yuds.
\end{aligned}
  \label{eq:utility_table}
\end{equation}
Typically $\yudh>\yuth$ and $\yuts>\yuds$, and $\yudh$, $\yuts$ are positive
and $\yuth$, $\yuds$ negative if we appropriately shift the zero of our
measurement units.

The expected utilities for dismissal and treatment are therefore
\begin{equation}
  \label{eq:expected_utilities}
    \expe{\yuD} = \yudh\, \ypth + \yuds\, \ypts,
\qquad
    \expe{\yuT} = \yuth\, \ypth + \yuts\, \ypts.
\end{equation}
Decision theory says the clinician ought to chose the action having maximum
expected utility. For example, she dismisses the patient if   $\expe{\yuD}
> \expe{\yuT}$, that is if
\begin{equation}
  \label{eq:dismiss_condition}
\ypts <  \frac{\yudh-\yuth}{\yudh-\yuth+\yuts-\yuds}.
\end{equation}

\section*{Conflict of Interest Statement}

The authors declare that the research was conducted in the absence of any
commercial or financial relationships that could be construed as a
potential conflict of interest.

\section*{Author Contributions}

CB constructed and analysed the graph data from fMRI scans. PGLPM developed
the statistical model. They both made the statistical analysis of the data
from the model. The manuscript was written PGLPM, CB, AM.

\section*{Funding}

We acknowledge partial support by the Helmholtz Alliance through the
Initiative and Networking Fund of the Helmholtz Association and the
Helmholtz Portfolio theme \enquote{Supercomputing and Modeling for 830 the
  Human Brain}.

\section*{Acknowledgments}

PGLPM thanks Mari \amp\ Miri for continuous encouragement
and affection; the kind staff at Iris, where part of this work was done;
Buster Keaton and Saitama for filling life with awe and inspiration; and
the developers and maintainers of \LaTeX, Emacs, AUC\TeX, Open Science
Framework, biorXiv, Hal archives, Python, Inkscape, Sci-Hub for making a
free and unfiltered scientific exchange possible. We thank Alper Yegenoglu
and Jakob Jordan for support and advice.

\bibliographystyle{frontiersinSCNS_ENG_HUMS}

\end{document}